\documentclass[lettersize,journal,10pt]{IEEEtran}

\usepackage{microtype}
\usepackage{subcaption}
\usepackage{graphicx}

\usepackage{booktabs}
\usepackage{adjustbox}
\usepackage{todonotes}
\usepackage{hyperref}
\usepackage{amsmath,amssymb,amsfonts}
\usepackage{stackengine}
\stackMath

\usepackage{multirow}
\usepackage{algorithm}
\usepackage{algorithmic}
\usepackage{makecell} 
\usepackage{float}
\usepackage{dblfloatfix}

\usepackage{cite}

\begin{document}
\bstctlcite{IEEEexample:BSTcontrol}

\title{KANtize: Exploring Low-bit Quantization of %\MT{b-spline-based?} 
Kolmogorov-Arnold Networks for Efficient Inference}

\author{Sohaib Errabii, Olivier Sentieys, Marcello Traiola\\Univ Rennes, Inria, CNRS, IRISA, Rennes, France}

\maketitle
\begin{abstract}
Kolmogorov-Arnold Networks (KANs) have garnered significant interest due to their potential for superior parameter efficiency and interpretability compared to Multi-Layer Perceptrons (MLPs). Their key innovation is the use of learnable non-linear activation functions, rather than traditional learnable linear weights.
%and predetermined non-linear activation functions.
Those functions are typically parametrized as spline functions, often expressed as linear combinations of basis splines (B-splines). The B-spline coefficients are the model's new learnable parameters.
However, spline function evaluation significantly increases inference computational complexity.
One conventional way to reduce such complexity is quantization, which reduces the numerical precision of parameters and activations. %, thereby lowering memory footprint and computational complexity.
The impact of quantization on KANs, and especially its effectiveness in reducing computational complexity, is largely unexplored, particularly for quantization levels below 8 bits.
%We need to more explicitly mention the goal: we open the KAN, look inside, and study the sensitivity to quantization of its different parts.
In this work, we examine KANs' behavior when using existing standard post-training weight and activation quantization and its impact on computational complexity and hardware efficiency.
%commonly employed in conventional Neural Networks.
%
%The study shows that different parts of KANs are affected differently by quantization, enabling different trade-offs between model accuracy and computational complexity.
% add that at 8 bit it works as in conventional NN (not kan)
%We find that quantization does not significantly reduce the accuracy of the B-spline recursive algorithm, while it substantially reduces its computational complexity and that of the overall KAN.
We find that B-splines are highly robust to quantization and yield substantial reductions in computational complexity, whereas the learnable parameters are more sensitive to quantization. 
Hence, we investigate the potential of using low-bit quantized precomputed tables as a replacement for the recursive B-spline algorithm. This approach aims to further reduce the computational complexity of KANs and enhance hardware efficiency while maintaining accuracy.
We performed experiments on six state-of-the-art KAN models, including both MLP-based and convolution-based models. Our results show that B-spline inputs/outputs can be quantized to 2-3 bits with negligible drop in accuracy, while substantially reducing computational complexity. 
For example, for ResKAN18, we show that a BitOps reduction of more than $50\times$ is possible without loss of accuracy, thanks to low-bit quantized B-spline tabulation.
Furthermore, using pre-computed 8-bit lookup tables enables up to $2.9\times$ inference speedup on GPUs, while on a systolic-array-based KAN accelerator, reducing the B-spline table precision from 8 to 3 bits on FPGA yields up to 36\% resource reduction, 50\% higher clock frequency and $1.24\times$ inference speedup. On a 28nm FD-SOI ASIC, reducing the B-spline bit-width from 16 to 3 bits achieves 72\% area reduction and 50\% higher maximum frequency.
%For example, for a ResNet18-based KAN model, only a $\textbf{1.8\%}$ drop in accuracy is observed on CIFAR-10 with 2-bit quantization, while the inference latency on the test set was reduced from 11 to 3.5 seconds, i.e., a 68\% improvement. 

\end{abstract}


\section{Introduction}

Kolmogorov-Arnold Networks (KANs) are a new Deep Neural Network (DNN) architecture originally designed as an alternative to Multi-Layer Perceptrons (MLPs)~\cite{kanpaper}.
KANs have gained significant attention and are now used in a range of applications, including time series analysis~\cite{vaca-rubio_kolmogorov-arnold_2024}, recommender systems~\cite{park_cfkan}, and medical image segmentation~\cite{li_u-kan_2024}. Their main advantages over traditional Deep Neural Networks (DNNs) are improved parameter efficiency and explainability.
The primary characteristic that makes KANs unique is the replacement of conventional edge weights with learnable activation functions. For example, learnable \textit{spline} functions have been used in the original paper~\cite{kanpaper}.
These splines $\phi$ can be expressed as a linear combination of basis functions (\textit{B-splines}) $b()$, i.e., $\phi(x)=\sum_{i} w_i b_{i}(x)$. 
Therefore, in KANs, the learnable parameters are the B-spline coefficients $w_i$.

However, from a computational perspective, adding spline function evaluation to the conventional multiplication significantly increases the computational cost, proportionally to the size of the basis. Indeed, to compute KAN inference, instead of performing a single scalar multiplication, each basis function must first be evaluated at the input, followed by a linear combination of all basis functions.
Moreover, their computation is costly because of the B-spline functions' recursive evaluation %using the Cox-de Boor formula
(see Section~\ref{sec:kan}).

Therefore, research efforts have been focusing on new approaches to accelerate KANs.
Among recent studies, compute-in-memory (CIM) approaches have been proposed~\cite{kacim,huang_hardware_2025}. In such studies, different ways to approximate the learned non-linear KAN functions or their basis are utilized, such as piece-wise linear (PWL)
approximation~\cite{kacim}.
A recent approach~\cite{arkane} focuses on the B-spline evaluation bottleneck. It proposes an efficient %systolic and wavefront 
dataflow acceleration methods for the Cox-de Boor recursive formula, achieving a significant speedup over CPU and GPU implementations.

Another conventional approach to tackling the high computing demands of DNNs is \textit{quantization}. The quantization process aims at reducing the bit precision of DNN weights and/or activations to low-bit representations (e.g., 8 bits or lower, instead of 32 bits) and to use simpler data types, e.g., fixed-point instead of floating-point. Not only does this reduce the DNN memory requirements, but it also enables the use of simpler and more efficient hardware (e.g., integer arithmetic units instead of floating-point ones) or packing more operands per instruction. In turn, this improves the latency and the overall efficiency.
A recent study had proposed a co-design strategy for efficiently implementing KANs on RRAM-based analog compute-in-memory (CiM) architectures~\cite{huang_hardware_2025}. In this context, a hardware-aware 8-bit quantization method was proposed to reduce the cost of B-spline evaluation, enabling an efficient 8-bit tabulation of B-splines.
More recently, a novel systolic-array-based accelerator that enables efficient end-to-end inference of KANs via non-recursive tabulated 8-bit B-spline evaluation was proposed, achieving a $2\times$ speedup over conventional systolic arrays~\cite{KANSAS}.
However, the quantization can be pushed to lower precision. %, with non-KAN conventional approaches using three~\cite{twn}, two~\cite{bnn}, and even one-bit~\cite{xnornet} quantization levels. 
A recent preprint studies KAN quantization, evaluating their accuracy loss when low-bit precision is used~\cite{QuantKAN}.
Unfortunately, such work neither studies computational complexity nor proposes methods to leverage quantization to achieve practical computational efficiency (e.g., faster inference, increased clock frequency, or optimized resource utilization). Moreover, it considers only the quantization of KAN's activations and coefficients, neglecting the quantization of \textit{B-spline} outputs, and does not explore the quantization space to assess the impact of quantizing the individual KAN tensor components and all their combinations. 
A recent paper explores the quantized tabulation of the entire set of KAN learnable \textit{spline} activation functions and their efficient mapping to FPGA LUT fabrics to reduce computational complexity, avoiding the B-spline basis computation~\cite{kanele}. The work highlights encouraging results with low precision (i.e., 5–8 bits). %However, no evaluation of precomputed bitwidth tables with fewer than 5-6 bits has been performed.
% 1. The impact of conventional quantization (i.e. W + A) on KAN is not explored - specify that we refer the MxM operands.
% 2. B-spline introduce 
%\MT{Here, maybe the text can say that currently available KAN-based models are very limited in the literature. If ConvKAN architectures exist, the text should cite them and, later in the exp section, state that we train our own models because no pre-trained ConvKAN models are available. Additionally, the text should note that no KAN-based Transformers or LLMs exist, and proposing such models is outside the scope of this study.}:
%Our results show that KANs can be quite resilient to quantization errors at small values of the grid size. Results on the convolutional KAN show only an accuracy drop of $2.9\%$ at $3$-bit weight and activation quantization on MNIST. The same model also shows an accuracy drop of $0.35\%$ when tabulating the B-spline with only $2$-bit addressing. The latter result is significant because the B-spline evaluation is the main bottleneck for inference (cf. Table~\ref{tab:forwardlatency}). The latter method exhibits similar performance on CIFAR-10, where a ResNet18-based model achieves a small accuracy drop of $1.8\%$ with only $2$ bits.

In this work, we explore the impact of standard post-training uniform integer quantization on KANs. Beyond conventional weight and activation quantization~\cite{QuantKAN}, we also consider quantizing B-spline outputs, which are specific to KAN inference and have not been explored previously. We study the impact of quantizing these three components (weights, activations, and B-spline outputs) both in isolation and jointly. Moreover, we study, for the first time, the beneficial effect of KAN's low-bit quantization on computational complexity by experimenting on GPU and a systolic-array-based KAN hardware accelerator~\cite{KANSAS}.
We show that the different KANs components exhibit distinct sensitivities to quantization. As also previously highlighted~\cite{QuantKAN}, we find that learnable parameters are particularly sensitive, with significant accuracy degradation below 5 bits for most models. Activations are more resilient than coefficients, tolerating 5-bit quantization with limited impact on accuracy. For the first time, we observe that B-spline outputs are the most resilient, tolerating quantization down to 3 bits for most models. %\MT{check this statement, is it valid for both inputs and outputs?} %\MT{x bits} %\MT{y\%} % <1% drop for 4/6 models at 3-bit quantB; CNN3 (3.57%) and ResKAN18 (1.73%) have higher drops We also perform a \textit{quantization space exploration} and show how different mixed-precision combinations offer different trade-offs between KAN accuracy and overall computational complexity.
%Results on the convolutional KAN show only an accuracy drop of $2.9\%$ at $3$-bit weight and activation quantization on MNIST. The same model also shows an accuracy drop of $0.35\%$ when tabulating the B-spline with only $2$-bit addressing. The latter result is significant because the B-spline evaluation is the main bottleneck for inference (cf. Table~\ref{tab:forwardlatency}). The latter method exhibits similar performance on CIFAR-10, where a ResNet18-based model achieves a small accuracy drop of $1.8\%$ with only $2$ bits.
%MT: put the general result about quantization e.g. "KANs still show good accuracy results with x-bit quantization, i.e., accuracy reduction less than y%.
%\todo[]{add summary of results}
Building on these findings, we explore the use of low-bit quantized precomputed tables to replace the recursive B-spline algorithm, significantly reducing KAN computational complexity while maintaining accuracy. While this has been explored for 8-bit quantization~\cite{KANSAS}, to the best of our knowledge, no evaluation of the computational advantages of low-bit quantized precomputed tables in KAN has been performed.
Our results show that, on a KAN-SAs-style systolic-array accelerator~\cite{KANSAS} deployed on FPGA, reducing the B-spline table precision from 8 to 3 bits lowers LUT utilization by 36\% at the same array size, enables a 50\% higher clock frequency, and yields a measured $1.43\times$ overall speedup while maintaining accuracy. ASIC synthesis in 28nm FD-SOI shows similar trends. Halving both B-spline and weight precision from 16 to 8 bits reduces cell area by 60\% and raises the maximum frequency by 33\%, while further reducing B-spline precision to 3 bits achieves an overall 72\% area reduction and 50\% higher frequency over the 16-bit baseline. Finally, on GPU, using 8-bit B-spline tabulation yields a $2.1$--$2.9\times$ inference speedup with no loss in accuracy.
Finally, we consider the use of low-bit quantized precomputed tables to replace the entire spline computation, as in~\cite{kanele}, and highlight the advantages and limitations of this approach compared to using precomputed tables to replace the recursive B-spline. Our exploration shows that, due to the nature of the two approaches, replacing spline computation is particularly convenient for models of limited dimension, whereas scalability becomes challenging for larger models.
In summary, the contributions of this paper are:
\begin{enumerate}
    \item A comprehensive evaluation of how a standard quantization scheme affects MLP-based and convolutional KANs. The sensitivity of the different tensor components of the KAN layers to quantization is investigated.
    \item An exploration of how combining different quantization levels across KAN tensor components yields different trade-offs between accuracy and computational complexity.
    \item An analysis of the improvements in inference runtime provided by \textit{B-spline} low-bit quantization and tabulation on GPU and a systolic-array-based KAN hardware accelerator, and the obtained hardware efficiency improvement when synthesizing the latter to FPGA and ASIC.
    \item An analysis of the trade-offs between accuracy and computational complexity achieved through low-bit quantization and tabulation of \textit{splines}, as well as its limitations when larger models are utilized.
\end{enumerate}

\section{Background and Motivation}

In this section, we summarize the theoretical background of Kolmogorov-Arnold Networks (Section~\ref{sec:kan}), which essentially involves replacing the weights of conventional neural networks with learnable splines.
In Section~\ref{sec:complexity}, we show the implications of this in terms of computational complexity and number of parameters.
Finally, in Sections~\ref{sec:quantization_background} and~\ref{sec:kan_approx}, we provide background on the methods used in this work, i.e., uniform integer quantization and tabulation of activation functions, respectively.

\subsection{Kolmogorov-Arnold Networks}\label{sec:kan}

The \textit{Kolmogorov-Arnold representation theorem} states that if $f$ is a multivariate continuous function on a bounded domain, then $f$ can be written as a finite composition of continuous functions of a single variable and the binary operation of addition \cite{KAtheorem, KAproof}. More specifically,
\begin{equation} \label{eq:1}
f(a)=f\left(a_1, \ldots, a_n\right)=\sum_{q=0}^{2 n} \Phi_q\left(\sum_{p=1}^n \phi_{q, p}\left(a_p\right)\right),
\end{equation}
where $\phi_{q,p}:[0,1] \rightarrow \mathbf{R}$ and $\Phi_q: \mathbf{R} \rightarrow \mathbf{R}$.

However, as there is no smoothness guarantee for these 1D functions, which is known to be critical for deep neural networks \cite{theoreticalissuesnn}, the Kolmogorov-Arnold representation theorem has not had much success in machine learning.
Recently, a generalization of this theorem defined a single layer that can be stacked to construct neural networks of arbitrary width and depth~\cite{kanpaper}.
Under this generalization, the Kolmogorov-Arnold representation theorem becomes a particular case of a 2-layer KAN $[n, 2n + 1, 1]$, i.e., the first layer has $N_{in} = n$ input neurons and $N_{out} = 2n + 1$ output neurons, and the second one $N_{in} = 2n + 1$, $N_{out} = 1$.

\begin{figure}[htbp]
\center
\includegraphics[trim={25 15 38 10},clip,width=0.5\textwidth]{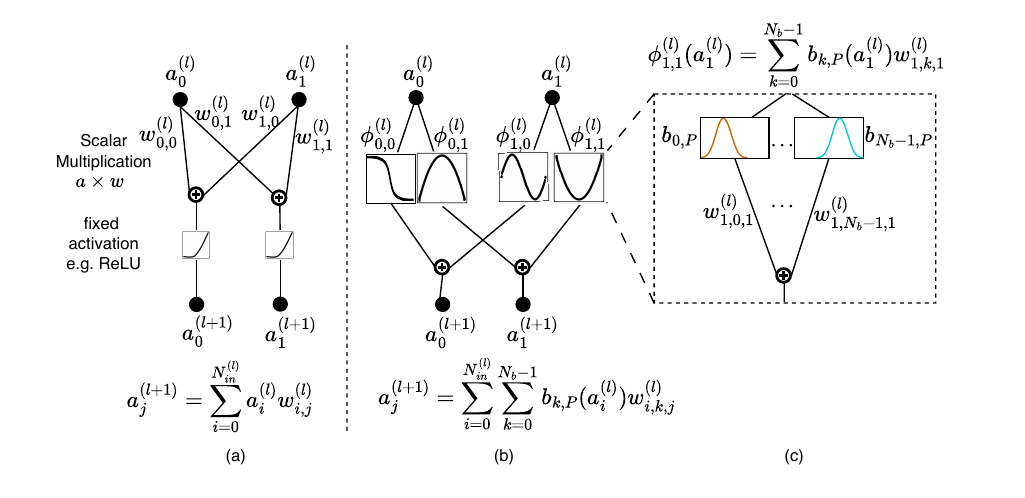}
\vspace{-0.4cm}
\caption{A single layer of dimensions [2, 2] for MLP (a) and KAN (b). Details of spline computation as a linear combination of B-splines}
\label{fig:kanvsmlp}
\end{figure}

%\MT{the $\sum_{p=1}^n \phi_{q, p}\left(x_p\right)$ contribution}Eq. \eqref{eq:1}
This generic KAN layer can be intuitively pictured as a 2-layer MLP (Figure~\ref{fig:kanvsmlp}(a)), where the weights at the edges are replaced with learnable activation functions, and the fixed activation (e.g., ReLU) at the nodes is simply replaced with a summation, as depicted in Figure~\ref{fig:kanvsmlp}(b).
%try to put figure as much as possible close to the point where you refer to them. Figures are a way to make the text more clear.
To address the missing guarantee of smoothness, the authors proposed to use parameterized splines as the learnable activations $\phi_{q,p}$. Splines are often used to interpolate or approximate data points in a smooth and continuous manner.
To parameterize these splines, various basis functions can be considered, such as Radial Basis Functions (RBF), Fourier, or B-splines \cite{kanbasisfuncs}. In this work, we focus on the B-spline basis functions (Figure~\ref{fig:kanvsmlp}(c)) originally proposed in \cite{kanpaper}, and widely used in KAN literature. Briefly, B-splines are evaluated as a function of the input (i.e., $b_{i,P}(a)$), the results are multiplied by learned coefficients $w_i$ and then summed to obtain the spline value $\phi(a)$.

\begin{figure}[htbp]
\centerline{\includegraphics[width=0.4\textwidth, trim={ 30 10 30 10}, clip]{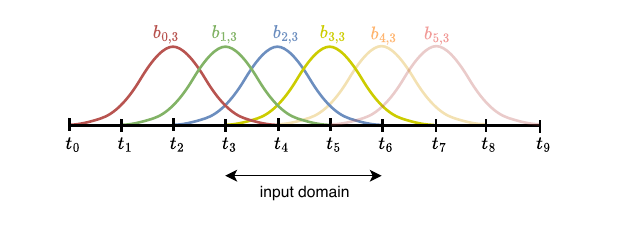}}
\caption{B-Spline basis for $G=3$, $P=3$ and $G+2P+1=10$ grid points $[t_0, \dots, t_9]$.}
\label{fig:bsplines}
\end{figure}

In more detail, B-spline is a family of piecewise polynomial functions $b_{i,P}$ defined by its degree $P$ (e.g., $P = 3$ for cubic 
splines as shown in Figure~\ref{fig:bsplines}) and a knot vector $t$, which specifies the points where the polynomial segments connect.
These functions can be computed by means of the Cox–de Boor recursion formula \cite{deboor} as
\begin{equation} \label{eq:deboor2}
b_{i, P}(a) = \frac{a-t_i}{t_{i + P} - t_i} b_{i, P-1}(a)+\frac{t_{i+P+1}-a}{t_{i+P+1}-t_{i+1}} b_{i+1, P-1}(a)
\end{equation}
with \begin{equation} \label{eq:deboor1}
b_{i, 0}(a) =
\begin{cases}
1 & \text { if } t_i \leq a<t_{i+1} \\ 0 & \text { otherwise. }
\end{cases}
\end{equation}
The suitability of B-splines as a basis for spline functions is due to a fundamental theorem from Curry and Schoenberg~\cite{curry1,curry2}, which states that any spline $\phi$ of degree $P$ can be expressed as a linear combination of B-splines, i.e., they form a basis for the spline space:
\begin{equation} \label{eq:spllin}
\phi_{i, j}(a)=\sum_{k} b_{k, P}(a)w_{i,k,j}
\end{equation}

B-spline basis functions have local support. Indeed, from the Cox-de Boor formula, we can derive that 
the function $b_{i, P}$ is only non-zero in the interval $[t_i, t_{i + P + 1}]$.
The grid $t$ is defined by first discretizing the input domain (the range of values taken by the KAN layer inputs) into $G$ intervals.
The grid is then extended by $P$ points on both sides in order to account for all non-zero B-splines within the input domain.
The grid has $G + 2P + 1$ points, $\left[t_{0}, t_{1}, t_{2}, \dots, t_{G + 2P}\right]$, within which $\textbf{G + P}$ non-zero B-splines are needed to evaluate the spline function according to Eq. \ref{eq:spllin}.
As an example, Figure~\ref{fig:bsplines} reports B-Spline basis functions of degree $P=3$. The input domain is divided into $G=3$ intervals, leading to $G+2P+1=10$ grid points, within which $G+P=6$ B-splines form a basis for the spline function evaluation. Each B-spline $b_{k, 3}$ is non-zero in the interval $[t_k, t_{k + 4}]$.

Finally, as reported in Eq.~\ref{eq:spllin} and sketched in Figure~\ref{fig:kanvsmlp}(b-c), evaluating $\phi_{i, j}(a_i)$ involves computing the $G + P$ B-splines at $a_i$, multiplying each by its corresponding learnable coefficient, and summing the results. Then, as indicated in Eq.~\ref{eq:kanlayereq}, all the spline functions are aggregated to produce the output of the layer.
\begin{equation} \label{eq:kanlayereq}
a^{(l + 1)}_{j} = \sum_{i=0}^{N^{(l)}_{in}}\sum_{k=0}^{N_b - 1}b_{k, P}(a^{(l)}_i)w^{(l)}_{i, k, j}
\end{equation}
This is equivalent to a matrix multiplication of the form,
\begin{equation}\label{eq:MM-formulation}
A^{(l + 1)} = B^{(l)}W^{(l)} = b(A^{(l)})W^{(l)}
\end{equation}
%Where, $B^{(l)}$ is the matrix of the evaluated B-spline functions at the layer inputs $a^{(l)}_{i}$.
%Where, $B^{(l)}$ is the matrix $(b_{k, P}(a^{(l)}_{i}))_{\substack{0 \le k < (G + P) \\ 0 \le i < N^{(l)}_{in}}}$
% In contrast to the MLP formulation,
% \begin{equation*}
% A^{(l + 1)} = \text{ReLU}(A^{(l)}W^{(l)})
% \end{equation*}

% \begin{figure}[htbp]
% \centerline{\includegraphics[width=0.4\textwidth]{figs/kanmatmulview.pdf}}
% \caption{The compute graph of a KAN layer $l$ consists of B-spline evaluation and matrix
% multiplication. The dimensions of the matrices $A^l$, $B^l$ and $W^l$ are respectively $M\times N^{(l)}_{in}$,
% $M\times N^{(l)}_{in}(G + P)$ and $N^{(l)}_{in}(G + P)\times N^{(l)}_{out}$. Where $M$ is the batch size and $N^{(l)}_{in}$, $N^{(l)}_{out}$ the number of input and output neurons of the layer $l$.}
% \label{fig:kanmatmulview}
% \end{figure}

%\paragraph*{\textbf{Matrix notations for layer of $N_l$ neurons}}

\noindent \textbf{Convolutional KAN:} The KAN operation can also be applied to convolution by replacing the scalar filter weights with learnable activations.
%\MT{discuss the main differences (if any) between what described before and ConvKANs. If there are none, explicitly say that the theory above applies to ConvKAN as well}
In practice, similarly to how standard convolution can be implemented as im2col followed by matrix multiplication~\cite{im2col}, a convolutional KAN unfolds the input feature map into a matrix of patches, then applies the same KAN operation.
An open-source implementation of this ConvKAN operation is available in~\cite{convkan}. Efficient training and design of CNN networks based on this convolution operation have also been studied in~\cite{torchconvkan}.
In this paper, to distinguish between the MLP- and CNN-based KANs, we refer to the convolutional one as \textit{ConvKAN}.

%\paragraph*{\textbf{KAN-based Transformers}} \MT{I think we can remove or put in the future works?}
%The advancement of KAN-based Transformers marks a promising new area of research. Recent studies have demonstrated that hybrids of KANs and Transformers can successfully substitute MLP blocks in Vision Transformers, thereby improving their interpretability and adaptability~\cite{TransUKAN,ViKANformer}. However, the lack of pre-trained models makes it challenging to evaluate KAN-based Transformer models such as Swin and ViT~\cite{KAN-survey}. Moreover, training B-spline-based KAN Transformers is computationally intensive, exhibiting higher complexity and slow convergence -- reaching almost 50 minutes per epoch, on the MNIST dataset on an NVIDIA A100 GPU~\cite{ViKANformer}. \MT{(Hence, we do not focus on KAN-based transformers in this study.) -- to change }
%Recent efforts have focused on KAN-based transformers with different basis functions, out of scope for this work. %~\cite{KAT}

\subsection{Increased number of operations in KANs}\label{sec:complexity} 

\begin{figure}[!t]
\centerline{\includegraphics[trim={31 17 48 14},clip,width=\columnwidth]{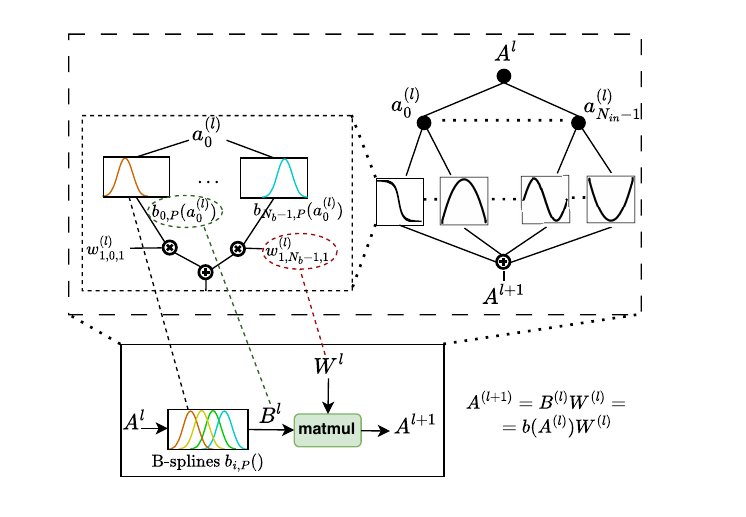}}
\caption{The compute graph of a KAN layer $l$ consists of B-spline evaluation followed by a standard matrix multiplication. %The dimensions of the matrices $A^l$, $B^l$ and $W^l$ are respectively $M\times N^{(l)}_{in}$, $M\times N^{(l)}_{in}(G + P)$ and $N^{(l)}_{in}(G + P)\times N^{(l)}_{out}$, where $M$ is the batch size and $N^{(l)}_{in}$, $N^{(l)}_{out}$ the number of input and output neurons of the layer $l$, respectively.
}
\label{fig:kanmatmulview}
\end{figure}

Figure~\ref{fig:kanmatmulview} sketches a KAN layer \textit{l} from a computational perspective. Firstly, B-splines are evaluated as a function of an activation matrix $A^l$, which generates an intermediate matrix $B^l$. The latter is then multiplied by the matrix of learned coefficients $W^l$.
The dimensions of the matrices $A^l$, $B^l$  and $W^l$ are respectively $M \times N_{in}^l$, $M \times N_{in}^l (G+P)$ and $N_{in}^l (G+P) \times N_{out}^l$, where $M$ is the batch size and $N_{in}^l$ and $N_{out}^l$  the number of input and output neurons of the layer \textit{l}.

%\MT{The text should make a clear link between the problem (this section) and which solution addresses which problem in later sections. Something like "we address this problem in section IIIA". As in Sec III the text reports first a solution to "Increased computation complexity" (i.e. b-spline tab+quant), maybe swapping the two problems in this section would help?}

\subsubsection{Increased model parameters and number of operations} 

Although KANs are more parameter efficient than the non-KAN counterpart~\cite{kanpaper}, parameterization using a function basis leads to an increase in model parameters by a factor of $G + P$, compared to non-KAN MLPs or CNNs, for the same layer dimension $\left[N^{l}_{in}, N^l_{out} \right]$.
For instance, a KAN layer with cubic B-splines ($P = 3$ and $G = 5$) would have $8\times$ as many parameters as an MLP layer of the same dimensions. 
This, in turn, increases the computational effort required. Indeed, the number of multiply-accumulate operations to perform the $B^{(l)}\times W^{(l)}$ matrix multiplication of Eq.~\ref{eq:MM-formulation} is $N^{(l)}_{in}\cdot(G + P)\cdot M \cdot
N^{(l)}_{out}$.

\begin{figure}[!b]
\centerline{\includegraphics[width=0.4\textwidth]{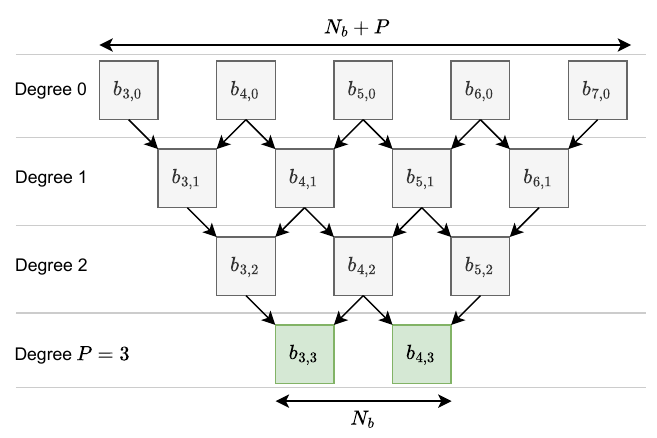}}
\caption{Cox-de Boor evaluation triangle for B-splines $b_{3,3}$ and $b_{4, 3}$ of degree $P = 3$.
Each row corresponds to the recursive computations of $b_{i, P}(x)$ from $b_{i, P - 1}(x)$ and $b_{i + 1, P - 1}(x)$.
The number of required B-splines in every row decreases linearly with $P$, leading to the total number of operations in Table~\ref{tab:complexity}.}
\label{fig:coxtriangle}
\end{figure}

For the B-spline evaluation ($B^{(l)}=b(A^{(l)})$), most KAN implementations rely on an iterative and parallel version of the Cox-de Boor formula of Eq.~\ref{eq:deboor2} to compute $N_b = G + P$ B-splines of degree $P$. Figure~\ref{fig:coxtriangle} conceptually illustrates it for only two B-splines, $b_{3,3}$ and $b_{4,3}$, of Figure~\ref{fig:bsplines}. 
As shown in the first row of the figure, the algorithm starts by evaluating the five B-splines of degree $0$ (cf. Eq.~\ref{eq:deboor1}) contributing to $b_{3,3}$ and $b_{4,3}$, then iteratively at each row $d$ it evaluates the B-splines of degree $d$ according to Eq.~\ref{eq:deboor1} until the row $d = P$. In the general form, it starts with $G + 2P$ B-splines of degree $d=0$. Then, iteratively at each row $d$, it evaluates the $G + 2P - d$ B-splines of degree
$d$ until the row $d = P$. 
Under the hypothesis that the reciprocals of the grid differences ($t_{i + P} - t_i$, $t_{i + P + 1} - t_{i + 1}$) are precomputed, four multiplications are needed per B-spline evaluation.
This is performed for the whole layer input matrix $A^{l}$ of size $M\times N^{(l)}_{in}$ leading to $4 MN^{(l)}_{in}(P(G + 2P) - \frac{P(P - 1)}{2})$ multiplications.
We report in Table~\ref{tab:complexity} the number of arithmetic multiplications (MULs) for an MLP layer versus a KAN layer.
When it comes to convolutional KANs, most implementations rely on lowering the convolution operation to a matrix multiplication using the im2col technique~\cite{chellapilla2006high}. 
Therefore, to obtain the arithmetic complexity of ConvKAN, we can replace $N_{out}$ and $N_{in}$ in Table~\ref{tab:complexity} respectively by $C_{out}$ and $K^2C_{in}H_{out}W_{out}$, with $K$ the filter size, $H_{out}, W_{out}$ the output image size, and $C_{in}, C_{out}$ the input and output number of channels.

From these considerations, we can estimate the computational complexity of KANs and explore the impact of quantization.
%The purpose of this work is to study the quantization of KANs. 
To do so, we estimate the number of operations using the conventional \textbf{BitOps} metric~\cite{zhou2018dorefanettraininglowbitwidth, li2020additivepowersoftwoquantizationefficient}. With such a metric, the cost of multiplying $n$-bit and $m$-bit integers can be approximated to $n\times m$ binary operations.
% From the number of multiplications reported in Table~\ref{tab:complexity}, and denoting by $bw_{A^{(l)}}$, $bw_{B^{(l)}}$ and $bw_{W^{(l)}}$ the bit-widths of activations $A^{(l)}$,  intermediate tensor of B-spline evaluations $B^{(l)}$ and weights $W^{(l)}$ respectively, the BitOps due to multiplications for the MLP layer is approximated as
\begin{table}[b]\caption{Number of multiplications (MULs) of $[N_{in}, N_{out}]$ MLP and KAN layers with batch size $M$.}
\centering
\setlength\tabcolsep{4pt}
\begin{tabular}{ccc}
\hline
& \makecell{MLP} & \makecell{KAN} \\
\hline
Matrix Multiply & $MN_{out}N_{in}$ & $MN_{out}N_{in}(G + P)$ \\
\hline
\makecell{Non Linearity \\ (B-splines)} & \textunderscore & $4 MN_{in}\left(P(G + 2P) -\frac{P(P - 1)}{2}\right)$ \\
\hline
\end{tabular}
\label{tab:complexity}
\end{table}
We define the bit-widths of the activations \(A^{(l)}\), the intermediate tensor of B-spline evaluations \(B^{(l)}\), and the weights \(W^{(l)}\) as \(bw_{A^{(l)}}\), \(bw_{B^{(l)}}\), and \(bw_{W^{(l)}}\), respectively. 
Then, based on the number of multiplications reported in Table~\ref{tab:complexity}, the multiplication BitOps for the a KAN layer $l$ can be approximated as
\begin{equation}\label{kanbitops}
\begin{split}
   \text{BitOps}^{(l)}_{KAN} &= MN_{out}^{l}N_{in}^{l}bw_{B^{(l)}}bw_{W^{(l)}} + \\
   &4\times MN_{in}^{l}\biggl[P(G + 2P) - \frac{P(P - 1)}{2}\biggr]bw_{A^{(l)}}^2 
\end{split}
\end{equation}
%\MT{not clear where the $^2$ comes from in (7), ie $bw^2_A$}
The last term in Eq.~\ref{kanbitops}, \(bw^2_{A^{(l)}}\), results from the Cox-de Boor recursion, where each multiplication has both operands involving activations ($a$), hence having the same bitwidth $bw_{A^{(l)}}$ (see Eq.~\ref{eq:deboor1}), unlike the matrix multiply involving B-spline and weights, hence having bit-width \(bw_{B^{(l)}}\) and \(bw_{W^{(l)}}\), respectively.

In comparison, the BitOps resulting from multiplications in the MLP layer does not include the additional term due to B-spline evaluation:
\begin{equation}\label{mlpbitops}
   \text{BitOps}^{(l)}_{MLP} = MN_{out}^{l}N_{in}^{l}bw_{A^{(l)}}bw_{W^{(l)}}
\end{equation}

Eq.~\ref{kanbitops} shows that the bit-width of the activations, $bw_{A^{(l)}}$, only impacts the arithmetic cost of the B-spline evaluation, while the bit-width of the weights $W^{(l)}$ and intermediate tensor $B^{(l)}$ (i.e., $bw_{W^{(l)}}$ and $bw_{B^{(l)}}$ respectively) contribute to the arithmetic cost of matrix multiplication.
In this paper, we study how quantizing weights, B-splines, and activations reduces BitOps and mitigates the intrinsic increase in model parameters and operations in KANs. In Section~\ref{sec:approach}, we further elaborate on how quantization reduces KAN computational complexity, and, building on this, how the tabulation of B-splines and learned splines further reduces it.

%For KANs, With typical values $G = 5$ and $P = 3$, the number of both multiplications and additions grows by an order of  magnitude compared to MLPs.
%For example, considering a layer with $N_{in} = 64, N_{out} = 10$, an MLP would require $640$ additions and multiplications. Meanwhile, a KAN with common parameters (i.e., $G =5, P = 3$) would require $23\times$ more additions and $18\times$ more multiplications.

%\begin{table}[ht]\caption{Arithmetic complexity of CNN and ConvKAN layers. $B = K^2 C_{in} H_{out} W_{out} C_{out}$}
%\centering
%\setlength\tabcolsep{3pt}
%\begin{tabular}{lll}
%\hline
%& \#Add. & \#Mul. \\
%\hline
%CNN & $B$ &  $B$ \\
%\hline
%ConvKAN & \begin{tabular}[c]{@{}l@{}} $B(G + P + 1)$ + \\ $C_{in}H_{in}W_{in}(G + P)\times$\\$(5P + 2)$\end{tabular} &
%\begin{tabular}[c]{@{}l@{}} $B(G + P)$ +\\ $C_{in}H_{in}\times$\\$W_{in}(G + P)4P$\end{tabular} \\
%\hline
%\end{tabular}
%\label{tab:convkan-complexity}
%\end{table}

\subsubsection{Computational overhead of B-spline evaluation} In addition to the increased number of operations,  the evaluation of the $G + P$ B-splines with the recursive Cox-de Boor formula is not accelerated efficiently on GPUs due to the interdependencies introduced by the recursion,  which intrinsically serializes the computation.
This directly impacts the inference latency. 
In particular, as further shown in Section~\ref{sec:results}, the B-spline evaluation completely dominates the latency of the forward pass. For instance, as reported in Table \ref{tab:forwardlatency}, the B-spline computation accounts for up to $\textbf{98\%}$ of the total inference time for MLP-based KANs, and $\textbf{78}$--$\textbf{84\%}$ for convolutional models where the im2col overhead reduces the B-spline contribution.
Therefore, for KANs, it is important to accelerate the evaluation of the basis functions~\cite{huang_hardware_2025,arkane,KANSAS}.
%only speeding up the linear layer's matrix multiplication would have little advantage. 
%Quantization techniques that enable the acceleration of matrix multiplication using specialized hardware, such as integer tensor cores, are less useful than speeding up the evaluation of the basis functions.
%
Similarly to what is shown in~\cite{huang_hardware_2025} and \cite{KANSAS} for a CiM-based and systolic-array-based accelerators, respectively, we show in Section~\ref{sec:bsplut} that this issue can be addressed in GPUs by replacing the recursive computations of B-splines with precomputed lookup tables (LUTs). This significantly reduces inference latency with negligible memory overhead and %an acceptable
accuracy drop, as shown in the results reported in Section~\ref{sec:results}. Moreover, we show that dedicated hardware~\cite{KANSAS} enables further improvements in KAN efficiency thanks to low-bit quantization.
Finally, %to jointly address the increased BitOps complexity and the inefficiency of the recursive algorithm, 
in Section~\ref{sec:spllut} we study the opportunities and challenges of a multiplier-free approach that uses low-bit-quantized precomputed tables to replace the entire \textit{spline} computation~\cite{kanele}. %This approach trades computational cost for memory usage and, as shown in Section~\ref {sec:results}, yields an acceptable degradation in accuracy. % in certain models.

%The next two sections provide background on the methods used in this work: uniform integer quantization and tabular optimization of activation functions.

\subsection{Uniform Integer Quantization}\label{sec:quantization_background}

Quantization has been extensively studied as an approach to reduce the memory footprint and inference latency of DNNs.
In this work, we focus on uniform integer quantization, which consists of mapping a floating-point range $[\alpha, \beta]$ to an integer range $[\alpha_q, \beta_q]$ that is representable by the considered precision (bit-width).

The linear transformation to map a real number $x$ to an integer $x_q$ is parameterized with  
$s \in \mathbf{R}$, $z \in \mathbf{Z}$ as follows:
\begin{align} \label{eq:h}
x & = \operatorname{dequantize}(x_q, s, z) = s(x_q - z)  \\
x_q & = \operatorname{quantize}(x, s, z, \alpha_q, \beta_q) \nonumber \\
& = \operatorname{clip}(\operatorname{round}\big(\frac{1}{s} x + z\big), \alpha_q, \beta_q)
\end{align}
where $\operatorname{clip}(x ; a, b) := \min (\max (x, a), b)$. 
The parameters $s, z$ are computed by mapping the floating-point bounds $[\alpha, \beta]$ to the quantization range bounds $[\alpha_q, \beta_q]$ as follows:
\begin{align}\label{eq:quant_params}
s & = \frac{\beta - \alpha}{\beta_q - \alpha_q} \\
z & = \operatorname{round}\left(\frac{\beta \alpha_q-\alpha \beta_q}{\beta-\alpha}\right) 
\end{align}

\subsection{Activation Function Optimization}\label{sec:kan_approx}

In digital neural networks, non-linear activations can be categorized into two main approaches from a computational perspective: (i) direct algorithmic implementation of the functions and (ii) use of precomputed Look-Up Tables (LUT) to replace the computation.
The first approach is generally implemented either by composing elementary functions (e.g., exponential, reciprocal, etc.) or by approximating them mathematically, for instance, with a polynomial.
This approach has been studied in various works, especially for architectures such as recurrent neural networks (RNNs) that require multiple evaluations of non-linear activation functions \cite{rnn_acts}. However, these works primarily focus on specific functions common in neural networks, such as the $tanh$ activation function \cite{tanh, highacctanh}.
In the second approach, precomputed activation function samples are stored in the LUT. The LUT is then accessed using the result of the previous layer \cite{dnnlut, 2foldlut}, and the corresponding precomputed output value is used.

As KANs rely even more heavily on activation functions than RNNs, these techniques become much more critical. Indeed, a feed-forward KAN layer consists of $N_{in}\cdot N_{out}$ spline functions, while ConvKAN has $K^2\cdot C_{in}\cdot C_{out}$ functions. In turn, as already mentioned, each spline is composed of $(G + P)$ B-splines.
In the context of the KAN activation functions, combining low-bit integer quantization and precomputed look-up tables has a twofold effect: (i) reducing the latency of computing the non-linear activation functions by reducing the amount of data to process and the computation complexity, and (ii) reducing the memory cost when using precomputed look-up tables for activation functions. The remainder of the paper presents the proposed approach for combining these two methods, aiming to achieve latency and memory improvements without compromising accuracy.

\section{KANtize Approach}\label{sec:approach}

%\MT{Here the text should highlight how the problems mentioned in Sec II are solved. The 'Increased computation complexity' issue is solved by tabulation (i.e. no recursion); the 'Increased model parameters and number of operations' problem is addressed by the approach 'Spline Tabulation and Quantization'. A clear link in the text will help the reviewers see how the contributions solve the problems. The text should also highlight upfront that a tradeoff exists in the second case (spline tab+quant) between memory and computation.}:

In this section, we explore the quantization and tabulation opportunities in KANs and combine them. 
%Figure~\ref{fig:kanmatmulview} illustrates this through the inference graph of two consecutive KAN layers.
%
Section~\ref{sec:wabptq} elaborates how quantization helps address the issue of increased model parameters and BitOps complexity. In particular, we discuss how quantizing the different KAN tensor components ($A^{(l)}$, $B^{(l)}$, and $W^{(l)}$) affects computational complexity.
%In particular, we first explore the effect of separately quantizing the weight matrix $W^{(l)}$, the activation matrix  $A^{(l)}$, and the intermediate tensor of B-spline evaluations $B^{(l)}$. We then exhaustively explore their combined quantization effects across 8-bit to 2-bit quantization.
%
Section~\ref{sec:bsplut} illustrates how to efficiently implement the B-spline $b()$ quantized tabulation scheme that helps address the inference bottleneck of the recursive \textit{B-spline} evaluation.
Finally, Section~\ref{sec:spllut} illustrates the tabulation of the learned \textit{splines} $\phi()$ (see Eq.~\ref{eq:spllin}), which completely replaces the B-spline evaluation and their linear combination depicted in Figure~\ref{fig:kanvsmlp}(c), and allows computing a KAN layer by visiting the tables and summing their values.

\subsection{Exploring the Quantization of KANs' Weights, Activations and B-splines}\label{sec:wabptq}

As shown in Figure \ref{fig:kanmatmulview}, from a computational perspective, weight quantization in a KAN layer is similar to MLP layers, with the coefficients of the basis functions acting as the weights (i.e., $W^{(l)}$).
Hence, when we apply conventional uniform integer weight quantization techniques to the KAN's coefficients, i.e., we reduce $bw_{W^{(l)}}$ in Eq.~\ref{kanbitops}, this significantly reduces the memory cost of weights and the BitOps complexity of the matrix multiply.
%
%While we only study simple min-max based quantization, we have also attempted GPTQ \cite{gptq}, a state-of-the-art 
%integer PTQ technique that finds the optimal quantization parameters by minimizing the reconstruction error of each 
%layer. However, we notice that the B-spline activation matrices are not full-rank which seems to be due (\MT{the text should not make vague statements (i.e., "IS due to")}) to the
%local support of B-splines (at most $P + 1$ values are non-zero for all inputs).
% The OBC~\cite{frantar2023optimalbraincompressionframework} framework was used when possible to quantize the weights of the models.
%
In MLPs, quantizing both weights $W^{(l)}$ and activations $A^{(l)}$, i.e., reducing $bw_{W^{(l)}}$ and $bw_{A^{(l)}}$ in Eq.~\ref{mlpbitops}, not only further reduces the memory cost and BitOps of the matrix multiplication but also enables more efficient hardware utilization, allowing the use of integer arithmetic pipelines or integer tensor cores. In KANs, since the B-spline outputs are used in the matrix multiplication, to achieve similar benefits we quantize the intermediate B-spline evaluation tensor, $B^{(l)}$, i.e., we reduce $bw_{B^{(l)}}$ in Eq.~\ref{kanbitops}. Finally, quantization of activations $A^{(l)}$ in KANs is helpful for reducing the memory and BitOps cost of computing the B-splines, i.e., reducing $bw_{A^{(l)}}$ in Eq.~\ref{kanbitops}.
% \MT{add an analysis on expected gains in different situations, e.g., in which cases quantizing A will help more than quantizing B and/or W according to equation 7}
As shown in Table~\ref{tab:complexity}, if $N_{out}$ is not too large, then the BitOps cost of B-spline evaluation can be as significant as the matrix multiplication. 
Therefore, %contrary to previous studies~\cite{QuantKAN}, 
to properly address the computational issues of KAN, it is important to consider quantizing all three $W^{(l)}$, $A^{(l)}$, and $B^{(l)}$ and investigating their combination. 
The results of this exploration are reported in Section~\ref{sec:quantization-results}.

\subsection{B-spline Tabulation}\label{sec:bsplut}

\begin{figure}[!b]
\center
\includegraphics[width=.9\columnwidth, trim={5 0 0 5},clip]{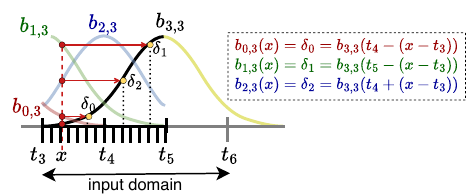}
\caption{Only half the values of one B-spline need to be stored (e.g., half of $b_{3,3}$, colored in black), and the other B-splines can be evaluated by translating the input and using symmetry. 
}
\label{fig:bspl_translation}
\end{figure}

\begin{figure}[!b]
\centering
\includegraphics[trim={4 5 10 10},clip,width=0.85\columnwidth]{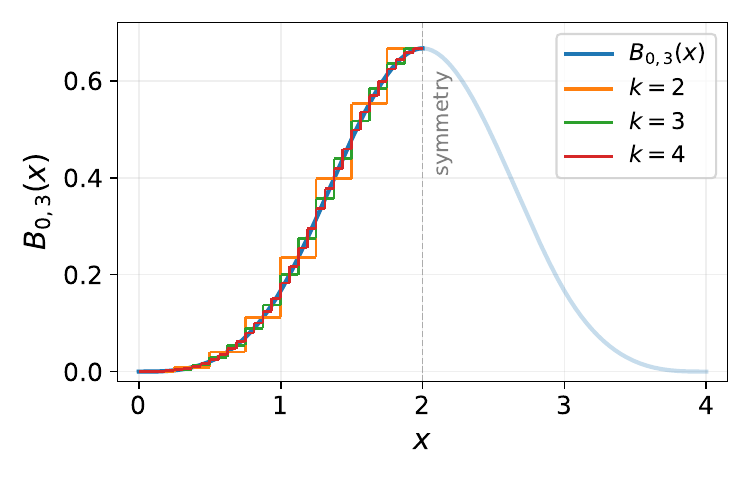}
\caption{Tabulation of the canonical B-spline $B_{0,3}(x)$: only the first half is stored with $2^k$ entries per knot interval $[t_i,t_{i+1}]$.}
\label{fig:bspl_table}
\end{figure}

\begin{figure}[!b]
\centering
\includegraphics[width=0.85\columnwidth]{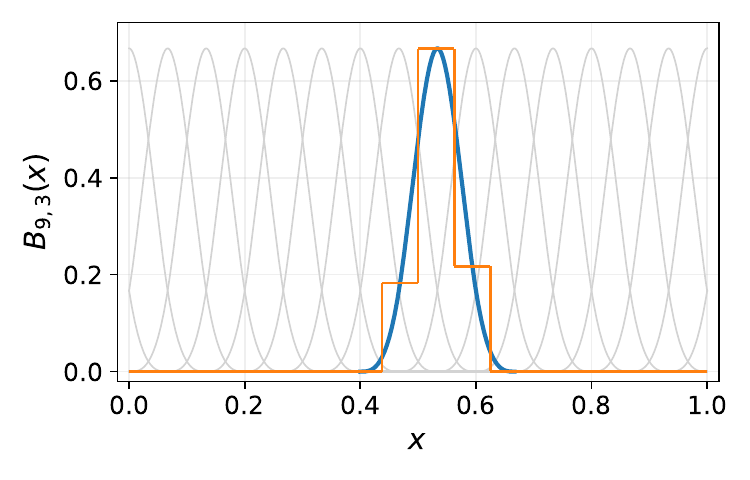}
\caption{Naive tabulation example with $G = 15$, $k = 4$: uniform quantization to $2^k$ levels wastes entries in zero regions, capturing only 5 of 16 entries for the B-spline.}
\label{fig:bspl_naive_table}
\end{figure}

As discussed in Section~\ref{sec:complexity}, the inefficiency of the recursive B-splines evaluation can be addressed by replacing the recursive computations with precomputed LUTs, significantly reducing inference latency.
As pointed out in previous literature~\cite{huang_hardware_2025,KANSAS,kanele} and depicted in the example in Figure~\ref{fig:bspl_translation} (adapted from Figure~\ref{fig:bsplines}), in the case of a uniform grid, we only need to tabulate half of a single canonical B-spline (highlighted in black in the figure). Then, we can infer the remaining elements from simple symmetry and translation.
For example, for a given input $x$, only the value $b_{3,3}(x)$ would be saved in the LUT, while the value $b_{1,3}(x)$ could be obtained by retrieving the value $b_{3,3}(t_5-(x-t_3))$, which is the same, as shown in Figure~\ref{fig:bspl_translation}.
% \MT{does this means training with a single G and then using different G values after training? This seems to be the case, as the next sentence talks about adapting G per layer during training}.
This enables a \textbf{single compact LUT} to serve all KAN layers, regardless of model depth, grid size, or input range.
When the grid is adapted to each layer's inputs during training~\cite{kanpaper}, the non-uniform knot spacing breaks translation invariance and would require separate tables.
However, techniques such as least-squares fitting enable approximating a general spline using a uniform B-spline basis, achieving arbitrary precision with a sufficiently fine grid.
Without loss of generality, this work focuses on uniform grids. A separate LUT would be needed only for each distinct spline degree $P$.

To build the LUT, the canonical B-spline is evaluated at equidistant points within each knot interval of the stored half-support, with boundary points mapped exactly to zero to avoid accumulated errors outside the 
local support.
As illustrated in Figure~\ref{fig:bspl_table}, when the tensor $A^{(l)}$ is quantized to $k = bw_{A^{(l)}}$~bits, the LUT contains $2^k$ entries per knot interval $[t_i,t_{i+1}]$.
For example, $k\!=\!2$ yields a coarse staircase of $4$ entries in each interval, e.g., [1,2], while $k\!=\!4$ gives $16$ entries and a much tighter approximation.
This contrasts with a naive strategy that uniformly quantizes the full input domain into $2^k$ levels.
As shown in Figure~\ref{fig:bspl_naive_table}, most entries would fall outside the B-spline's narrow local support, wasting storage and yielding a poor approximation, especially at higher~$G$.

The values stored in the LUT can also be quantized to $h = bw_{B^{(l)}}$~bits, corresponding to the quantization of tensor $B^{(l)}$. We refer to the resulting configuration as a $k{\times}h$ LUT.
This notation denotes the quantization bit-widths, not the table dimensions. Since the half-support spans $\lceil(P+1)/2\rceil$ knot intervals with $2^k$ entries of $h$~bits each, the actual LUT memory is $2^k \times \lceil(P+1)/2\rceil \times h$~bits.

As highlighted in previous research~\cite{KANSAS,kanele}, tabulation further reduces the computational complexity, as it replaces the cost of recursively computing the B-spline values with the cost of fetching data from a LUT. Also, it ensures that the LUT only stores the non-zero B-spline values. However, without specialized hardware support for efficient sparsity management (e.g., as in~\cite{KANSAS}), reconstructing the complete sparse matrix is necessary to perform matrix multiplication with the coefficients. 

We study, for the first time, the effects of tabulating the B-splines with different $k$ and $h$ low-bit values in terms of the trade-off between accuracy and memory needed to store them. Furthermore, we evaluate the gains in terms of BitOps reduction and how they translate to inference latency reduction. The inference latency gains, of course, depend on the HW support. For instance, GPUs only support specific bit-widths natively (e.g., 8 bits), and would not gain in latency when using lower-bit quantizations (e.g., less than 8 bits). However, using custom hardware allows further improvement in latency when using lower-bit quantizations. The results of the study are presented in Section~\ref{sec:results-B-spline}.

\subsection{Spline Tabulation and Quantization}\label{sec:spllut}

Finally, we explore the impact of low-bit quantization on a multiplier-free KAN inference.
Instead of obtaining the learned splines through linear decomposition in the B-spline basis followed by matrix multiplication (as shown in Figure~\ref{fig:kanmatmulview}), we directly tabulate each learned function independently on the
extended grid domain, as also shown in~\cite{kanele}. % for 8-bit quantized values.
Thus, given a layer of dimensions $\left[N_{in}, N_{out} \right]$, the only operations remaining after tabulation of the splines are
the $N_{in} \cdot N_{out}$ additions that sum the values retrieved from the corresponding $N_{in} \cdot N_{out}$ spline tables.
\begin{figure}[b]
  \vspace{-10pt}
  \centering
  \includegraphics[trim={25 15 80 10},clip,width=.9\columnwidth]{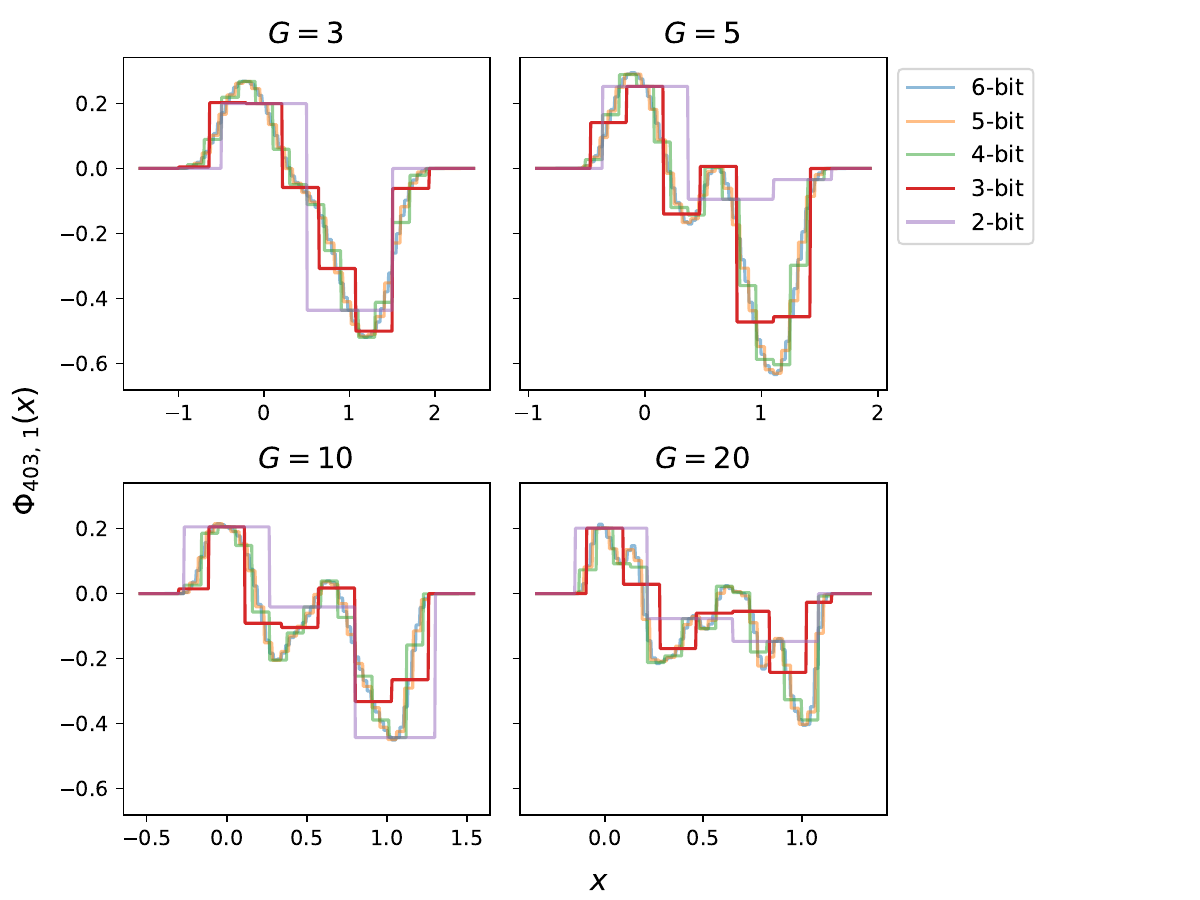}
  \caption{Example of a learned spline in the single-layer KAN $[784, 10]$ for different $G$ values. The legend shows the number of bits used for input quantization.}
  \label{fig:spline_table_single_layer}
\end{figure}
Figure~\ref{fig:spline_table_single_layer} shows examples of different splines tabulated by sampling the inputs with different bit-widths and for different $G$ values. For example, when inputs are quantized with $bw_{A^{(l)}}\!=\!3$ bits, the resulting table has $2^3=8$ entries, each storing the spline output value at the corresponding quantized input level (red curve).
Similar to B-spline tabulation, the values stored in the spline tables can also be quantized to 
For comparison, the original FP32 coefficients require $(G+P) \cdot 32$ bits per connection.
further reduce memory cost. Indeed, the memory cost is significant in this case because the table dimension scales with the layer dimensions. A single KAN layer leads to $N_{in} \cdot N_{out} \cdot 2^{(bw_{A^{(l)}})} \cdot h$ stored bits when using $h$-bit quantized spline values and $2^{(bw_{A^{(l)}})}$ table entries.
For configurations where $2^{(bw_{A^{(l)}})} \cdot h < (G+P) \cdot 32$,
the spline table would actually be smaller than the FP32 $W^{(l)}$ coefficients it replaces.

%The quantization parameters can be computed without any calibration. This is a benefit from the local support of B-splines. Since all B-splines are zero outside of the grid range (see Figure \ref{fig:bsplines}), then the splines, as their linear combination, are also all zero outside of the grid. This range can thus be used to compute the quantization parameters.
The quantization parameters can be computed without requiring any calibration, which is a direct benefit of the local support 
property of B-splines. Since all individual B-spline functions are zero outside their respective grid intervals (as 
illustrated in Figure \ref{fig:bsplines}), any linear combination of these splines will also be zero outside the overall grid 
range. This characteristic enables us to utilize the defined grid range for computing the quantization parameters, thereby 
simplifying the process by eliminating the need for calibration data.
For example, as shown in Figure~\ref{fig:spline_table_single_layer}, all splines of a given layer share the same grid and are zero outside the extended grid range. Therefore, the quantization range for the input can be set directly to the grid bounds, since any activation falling outside this range would have zero contribution regardless of its actual value.
Furthermore, the same grid is often used across the KAN layers. Thus, the same quantization parameters are used to map the previous layer's output to an integer for addressing the spline tables.

\textbf{Scalability of spline tabulation}:
While spline tabulation eliminates multiplications, the total number of spline tables scales with $\sum_l N_{in}^{(l)} \cdot N_{out}^{(l)}$, which grows quickly with model size. The implications of this scaling for both GPU inference latency and FPGA resource cost are studied in Section~\ref{sec:results_spline_latency} and Section~\ref{sec:results_spline_scalability}, respectively.

\section{Experimental Results}
\label{sec:results}

% \MT{to be revised: Unlike well-established CNN architectures, KAN-based architectures remain scarce, and
% mostly consist of adapting previous architectures by replacing the linear or convolution 
% operations. To this date, there is no official pre-trained KAN model in popular frameworks 
% such as PyTorch. Moreover, the few pre-trained KAN models that we have managed to test exhibit collapsed coefficients, i.e., all the coefficients of the basis functions are zero.
% Therefore, these models rely predominantly on the bias branch that consists of a Sigmoid Linear Unit (SiLU) followed
% by a linear transformation, effectively reducing the KAN layer to an MLP layer.
% As a result, to study the post-training quantization dynamics of the KAN layer, we trained
% our own CNN and MLP KAN models without the SiLU bias.}

%MT: to reintegrate when Transformer experiments are available
% The KAN literature to date focuses on feed-forward and convolutional (ConvKAN) designs, standing apart from the 
% state of the art models such as transformers. The first reason is the mentioned parameter growth when adapting 
% conventional architectures to use KAN layers, and second, the recursive evaluation of B-splines dominates the 
% compute latency. Thereby making the training and experimentation of large KANs significantly more challenging.  

In line with previous work on KANs~\cite{kanpaper,QuantKAN,kanele,KANSAS}, in our evaluation we focus on the two state-of-the-art KAN families, MLP-based and ConvKAN, that we can feasibly train for classification on MNIST and CIFAR-10, given the lack of open-source pretrained KAN models.
%
% The MLP-based KAN which we denote by their layer widths (e.g. $[784, 64, 10]$ for two consecutive layers $[784, 
% 64]$ and $[64, 10]$) that we train on MNIST. And CNN-based KANs  LeKAN for the model based on 
% LeNet~\ref{fig:lekan_arch} which was also trained on MNIST. CNN3 and CNN4 are slightly larger ConvKANs which simply consist of three and four consecutive convolution layers respectively, trained on both MNIST and CIFAR-10. And finally ResKAN18 obtained by swapping in the ConvKAN operation instead of the conventional convolution.
%
For MNIST classification, we use MLP-based KANs and a ConvKAN model based on LeNet~\cite{lenet} with two ConvKAN layers followed by a single feed-forward KAN layer. %, as illustrated in Figure \ref{fig:lekan_arch}.
For \mbox{CIFAR-10}, we study two simple convolutional models, CNN3 and CNN4, with three and four ConvKAN layers, respectively, and a ResNet18-based model in which convolutions are replaced by ConvKAN layers and thus called ResKAN18.
Table~\ref{tab:models} summarizes the models, datasets, and number of parameters.

\begin{table}[b]
\caption{Models used in the evaluation ($P\!=\!3$, $G\!=\!3$). Dimensions denote layer widths for KAN models and channel count for ConvKAN models ($3{\times}3$ kernels unless noted).}
\centering
\begin{tabular}{l l l l r}
\hline
Model & Type & Dataset & Dimensions & Params \\
\hline
KANMLP1 & KAN & MNIST & $[784, 10]$ & 47\,K \\
KANMLP2 & KAN & MNIST & $[784, 64, 10]$ & 305\,K \\
LeKAN & ConvKAN & MNIST & $[1, 6, 16]$ ($5{\times}5$) & 39\,K \\
CNN3 & ConvKAN & CIFAR-10 & $[3, 32, 64, 128]$ & 560\,K \\
CNN4 & ConvKAN & CIFAR-10 & $[3, 32, 64, 128, 512]$ & 4.1\,M \\
ResKAN18 & ConvKAN & CIFAR-10 & ResNet18 & 67\,M \\
\hline
\end{tabular}
\label{tab:models}
\end{table}

% \begin{figure}[b]
% \centerline{\includegraphics[width=0.5\textwidth]{figs/lekan.pdf}}
% \caption{LeKAN: LeNet-based convolutional KAN for MNIST classification \cite{nnsvg}}
% \label{fig:lekan_arch}
% \end{figure}

For all models, we consider that all KAN layers in a given model have the same $G$ and $P$ parameters and a uniform grid that is not updated during training. Unless otherwise specified, all models were trained without using the bias based on SiLU, suggested by \cite{kanpaper}. 
This bias is, in fact, an independent MLP branch, and thus significantly impacts the training and the resulting weight distribution between actual basis function coefficients and the scaling used for SiLU. In this paper, we focus on studying the behavior of the spline-based KAN operation itself. 

% We limit our experiments to MNIST and CIFAR-10, due to the lack of public pre-trained KAN models on large datasets 
% such as ImageNet. \MT{This text should be moved at the beginning (cf. comment at the beginning of this section) }

In this section, we report the results of the described quantization and tabulation approaches on the models.
Section~\ref{sec:quantization-results} reports the results of quantization in terms of the trade-off between accuracy and BitOps reduction.
Section~\ref{sec:results-B-spline} studies the trade-off between accuracy and memory reduction thanks to B-spline tabulation, and discusses the results concerning the improvements in inference latency obtained through tabulation.
Finally, Section~\ref{sec:spline_tabul_results} examines these same aspects for spline tabulation and the associated scalability challenges.

\subsection{Quantization Space Exploration: Accuracy vs BitOps}
\label{sec:quantization-results}

\begin{figure*}[htbp] 
    \centering
    % \begin{subfigure}[b]{\textwidth}
    %     \includegraphics[width=\textwidth]{figs/acc_vs_bops/single-quantization-accVSBitops.pdf}
    %     \caption{Accuracy vs BitOps results when quantizing only one component of the KAN layer at a time, i.e., $W^{(l)}$, $A^{(l)}$, and $B^{(l)}$.}
    %     \label{fig:single-quantization-accVSBitops}
    % \end{subfigure}
    \begin{subfigure}[c]{0.55\textwidth}
       \includegraphics[width=\textwidth]{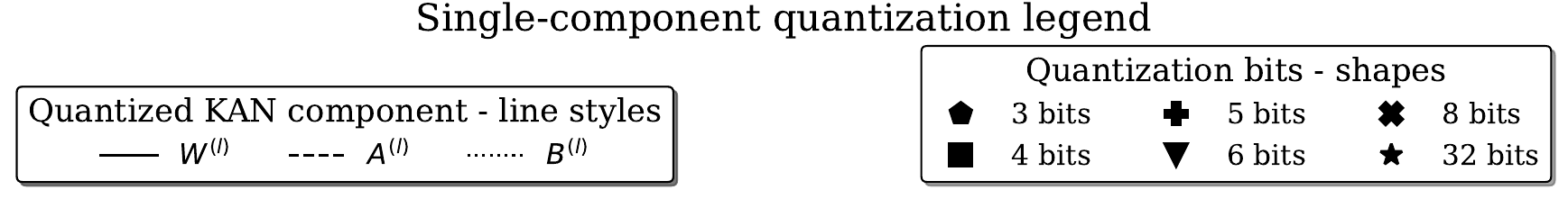}
    \end{subfigure}\\
    \begin{subfigure}[b]{0.32\textwidth}
        \includegraphics[width=\textwidth]{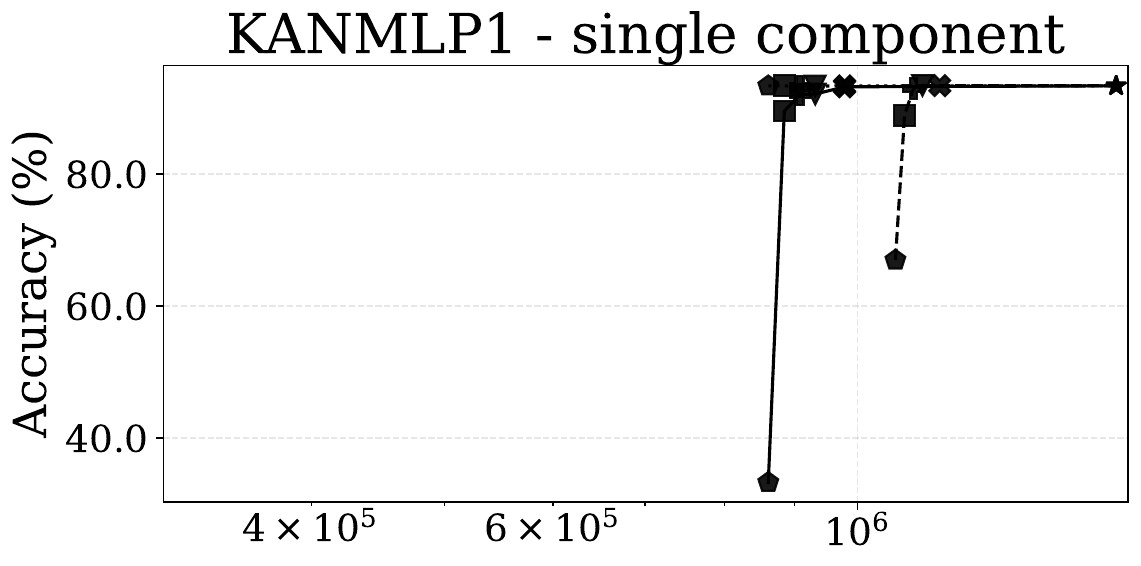}
        \vskip -7pt \caption{}
    \end{subfigure}
    \begin{subfigure}[b]{0.32\textwidth}
        \includegraphics[width=\textwidth]{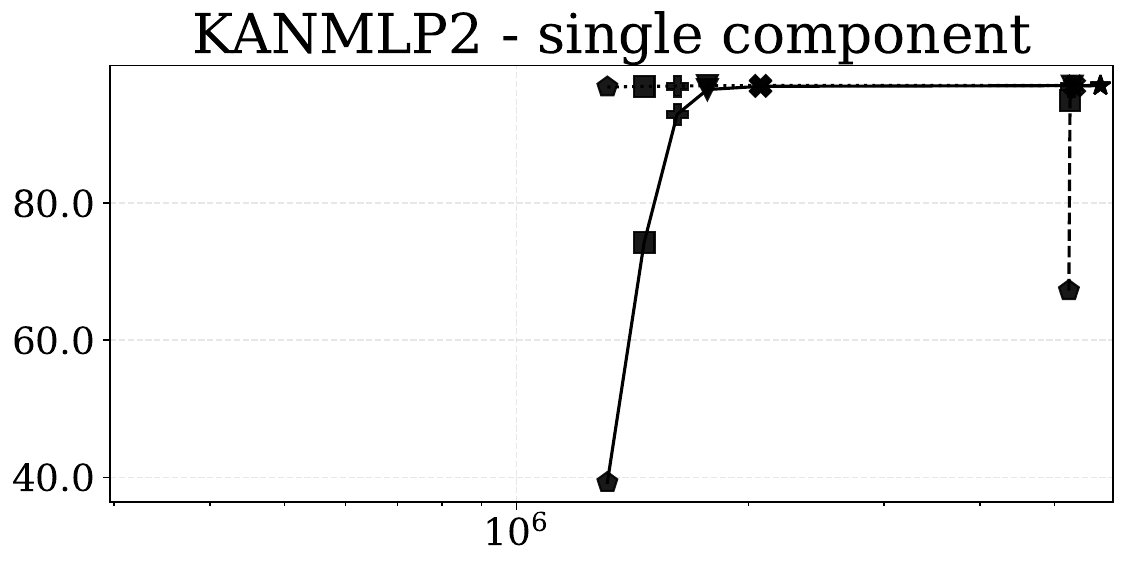}
        \vskip -7pt \caption{}
    \end{subfigure}
    \begin{subfigure}[b]{0.32\textwidth}
        \includegraphics[width=\textwidth]{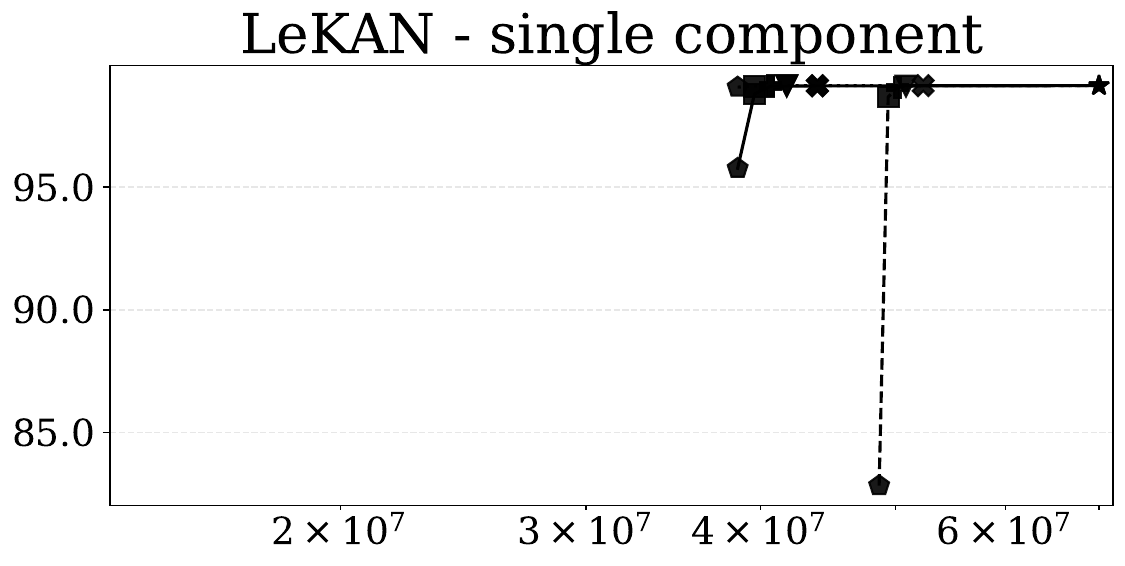}
        \vskip -7pt \caption{}
    \end{subfigure}
    \vspace{-2pt}
    \begin{subfigure}[b]{0.32\textwidth}
        \includegraphics[width=\textwidth]{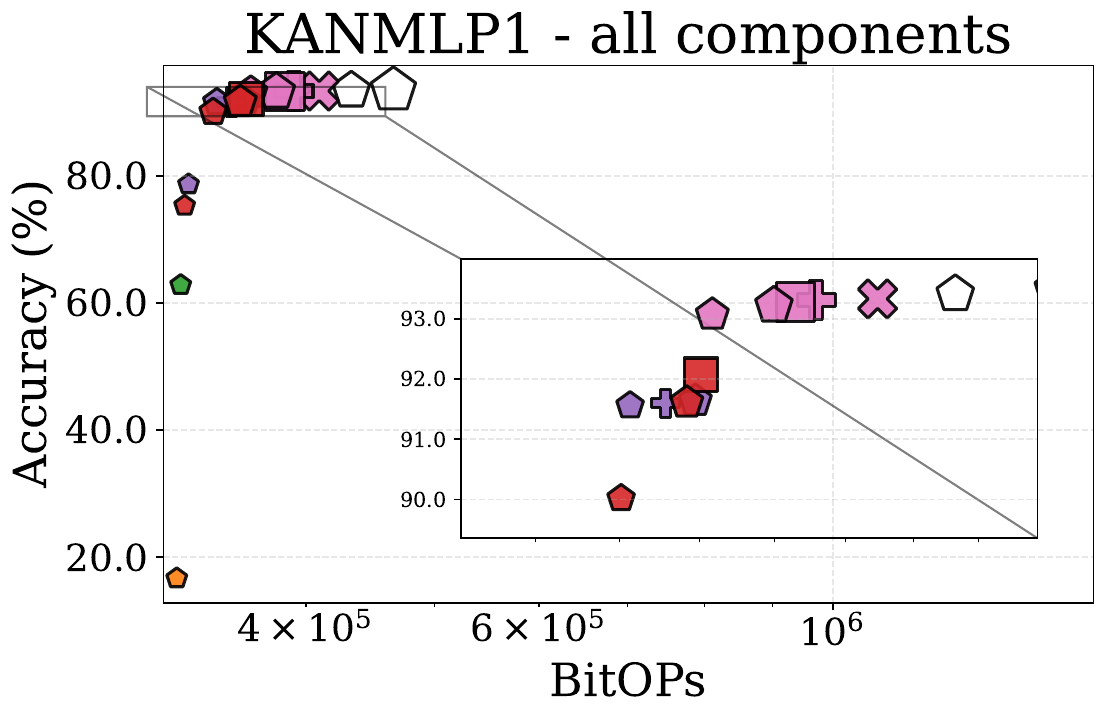}
        \vskip -7pt \caption{}
    \end{subfigure}
    \begin{subfigure}[b]{0.32\textwidth}
        \includegraphics[width=\textwidth]{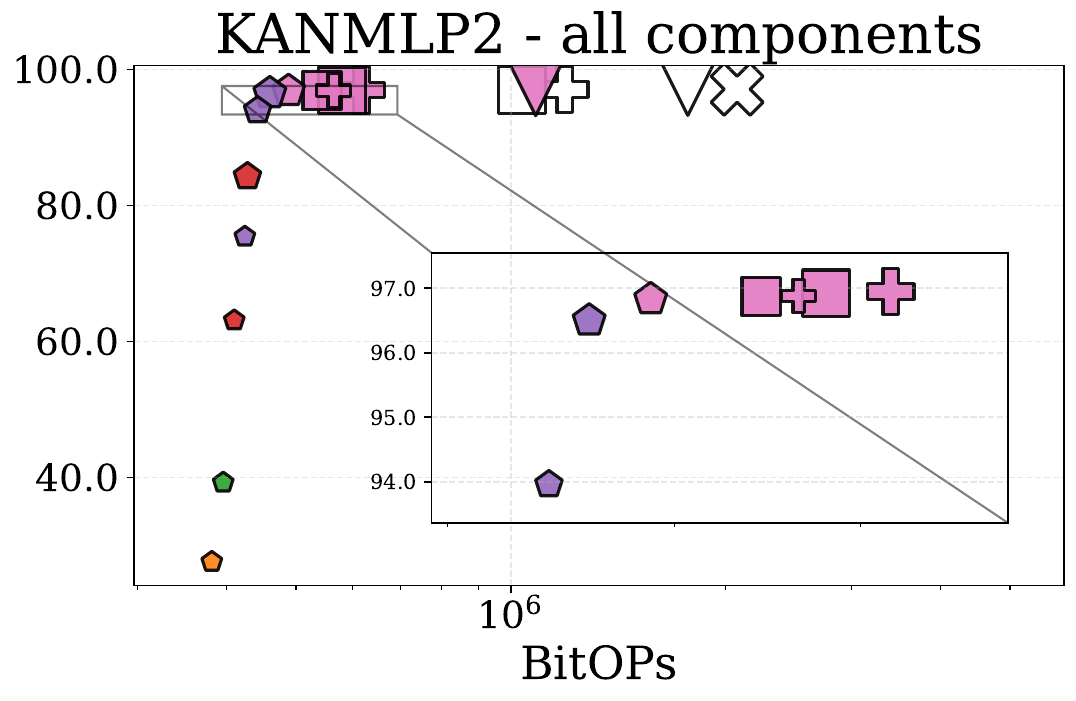}
        \vskip -7pt \caption{}
    \end{subfigure}
    \begin{subfigure}[b]{0.32\textwidth}
        \includegraphics[width=\textwidth]{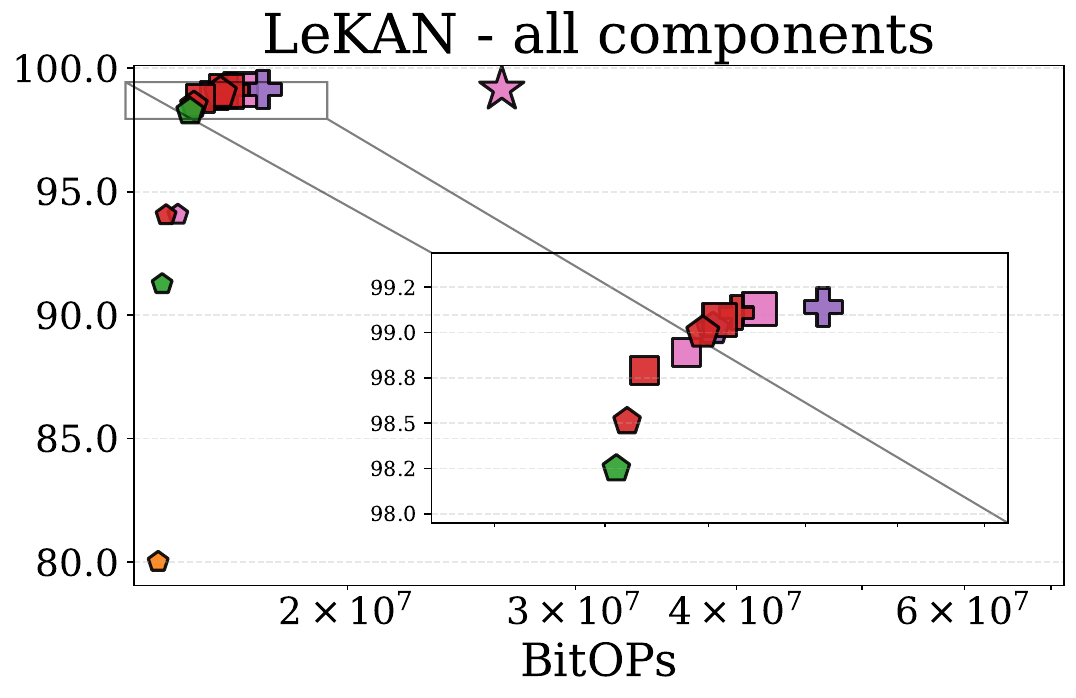}
        \vskip -7pt \caption{}
    \end{subfigure}
    \begin{subfigure}[c]{0.85\textwidth}
        \includegraphics[width=\textwidth]{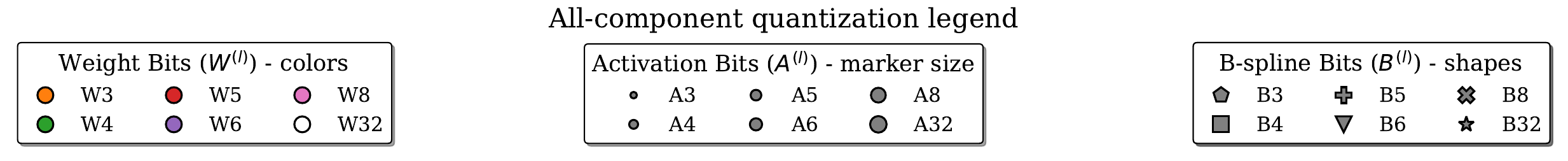}
\end{subfigure}\par\vspace{-.1\ht\strutbox}\noindent\hrulefill\par
        \begin{subfigure}[b]{0.32\textwidth}
        \includegraphics[width=\textwidth]{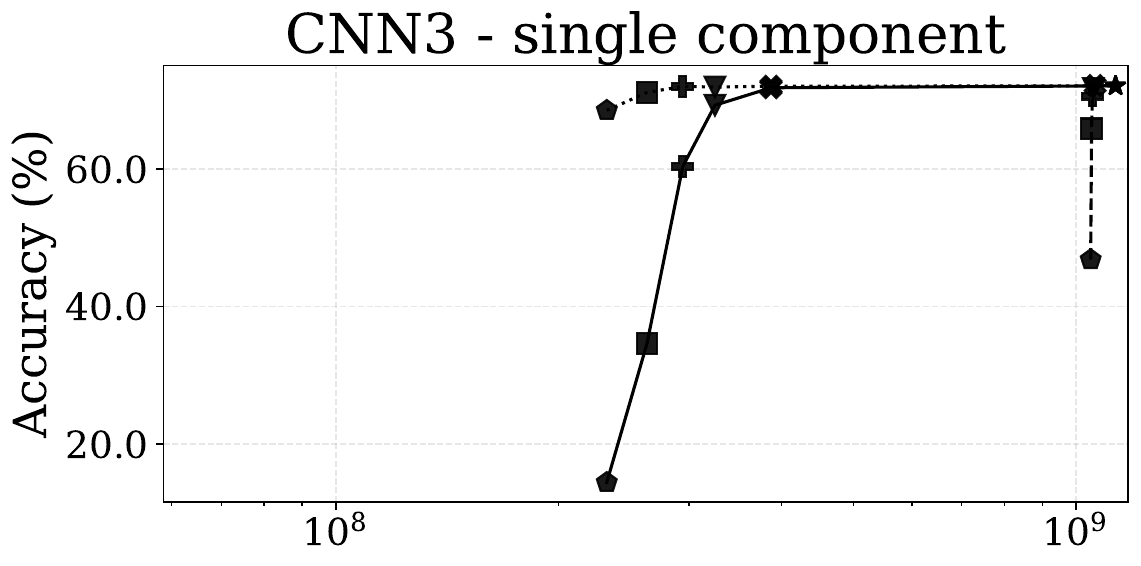}
        \vskip -7pt \caption{}
    \end{subfigure}
    \begin{subfigure}[b]{0.32\textwidth}
        \includegraphics[width=\textwidth]{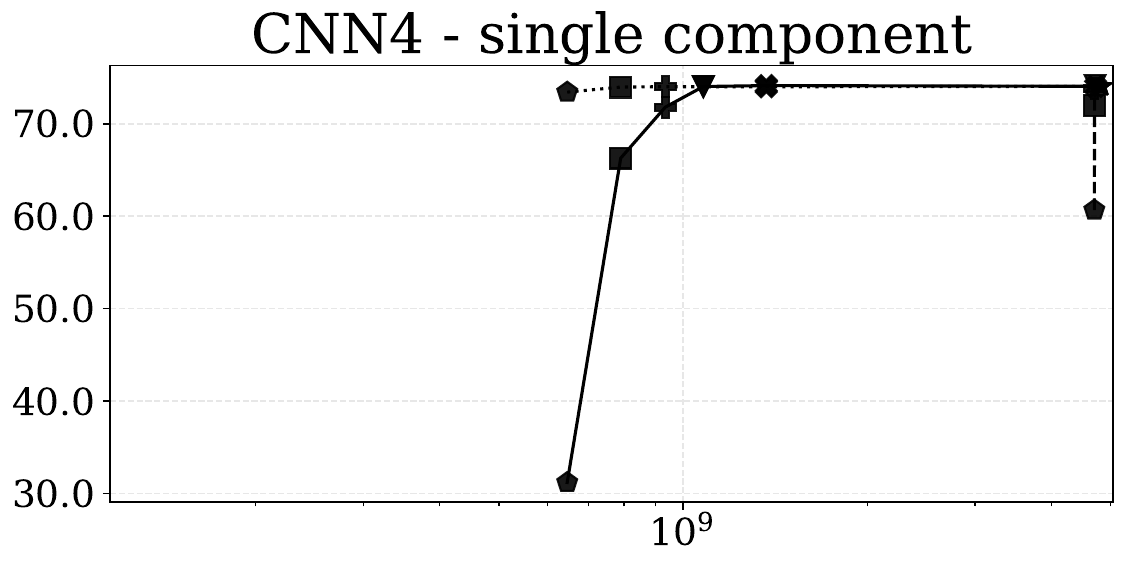}
        \vskip -7pt \caption{}
    \end{subfigure}
    \begin{subfigure}[b]{0.32\textwidth}
        \includegraphics[width=\textwidth]{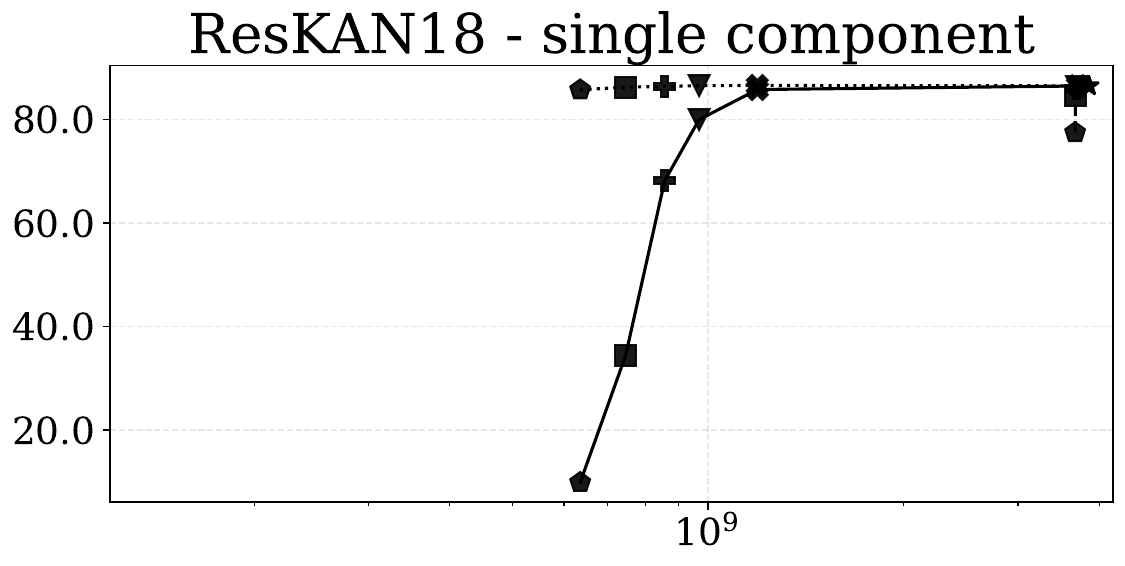}
        \vskip -7pt \caption{}
    \end{subfigure}
    \vspace{-2pt}
    \begin{subfigure}[b]{0.32\textwidth}
        \includegraphics[width=\textwidth]{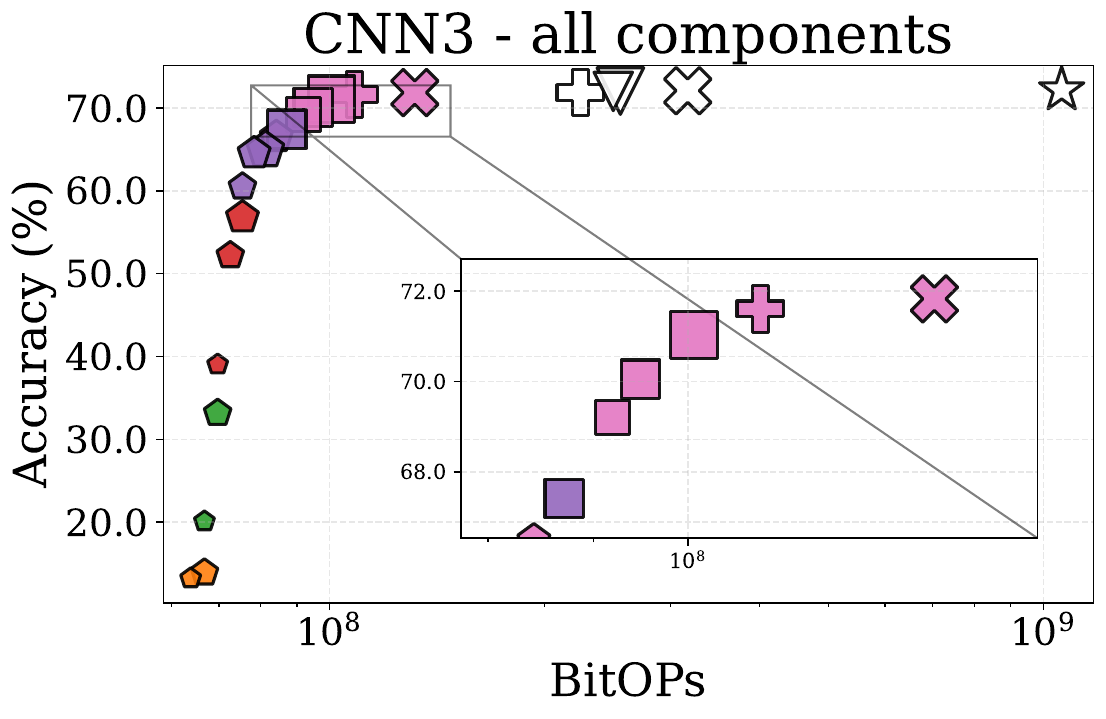}
        \vskip -7pt \caption{}
    \end{subfigure}
    \begin{subfigure}[b]{0.32\textwidth}
        \includegraphics[width=\textwidth]{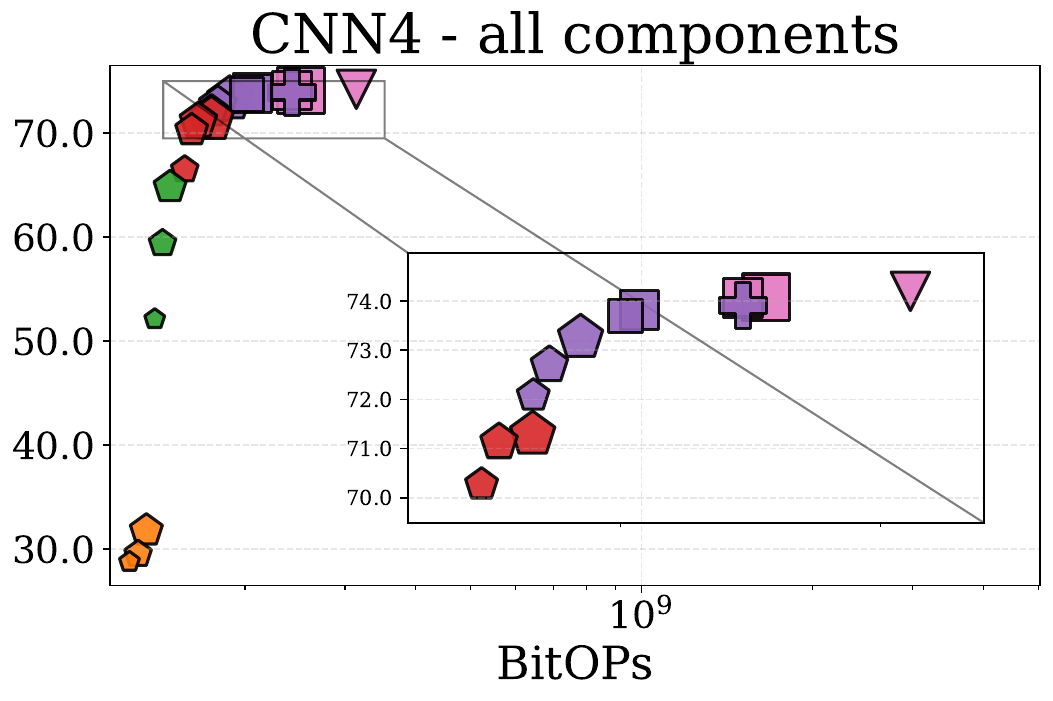}
        \vskip -7pt \caption{}
    \end{subfigure}
    \begin{subfigure}[b]{0.32\textwidth}
        \includegraphics[width=\textwidth]{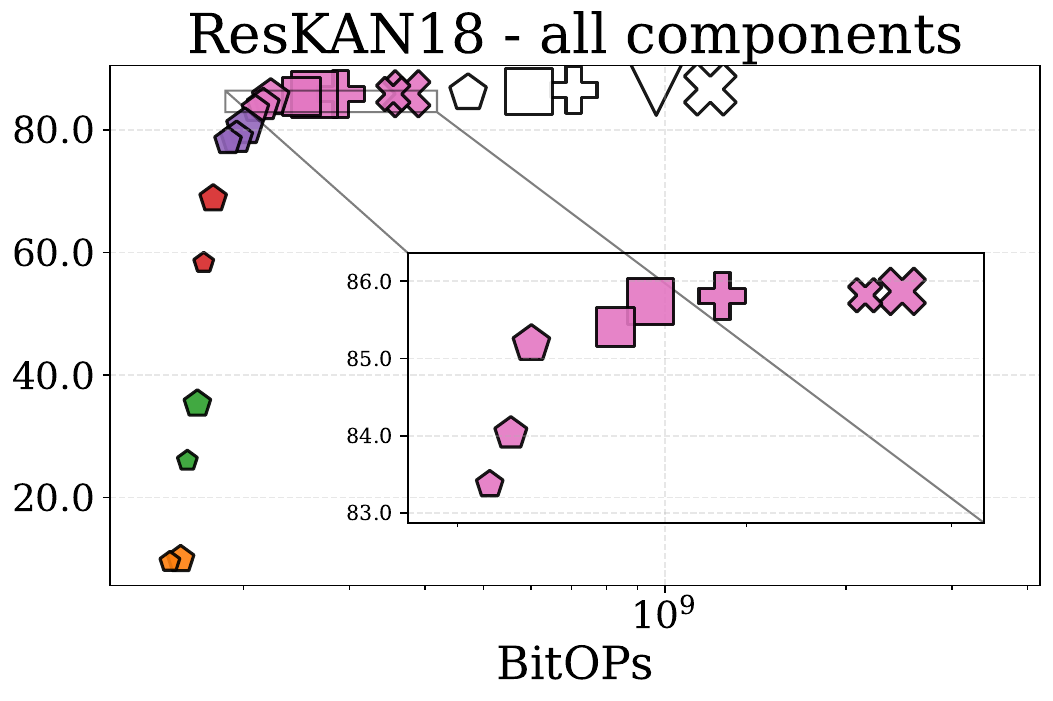}
        \vskip -7pt \caption{}
    \end{subfigure}    
\caption{Accuracy vs BitOps trade-off results of KAN quantization. \textit{Single component} graphs (a,b,c,g,h,i) report results when quantizing only one component of the KAN layer at a time, i.e., $W^{(l)}$, $A^{(l)}$, and $B^{(l)}$. For different components, different line styles are used, and for different bit widths, different shapes. \textit{All component} graphs (d,e,f,j,k,l) show the Pareto front of results obtained through quantizing multiple components of the KAN layer at the same time. The bit-widths of $W^{(l)}$, $A^{(l)}$, and $B^{(l)}$ are encoded by marker colors, sizes, and shapes, respectively.}
\label{fig:group-quantization-accVSBitops}
\end{figure*}

As mentioned in Sections~\ref{sec:complexity} and~\ref{sec:wabptq}, the quantization of the different tensor components of the KAN layer $W^{(l)}$, $A^{(l)}$, and $B^{(l)}$ impacts the computational cost (measured in BitOps using Eq.~\ref{kanbitops}) and the model's accuracy differently.
To isolate the effect of each, we first quantize only one component at a time 
%to specific bit-widths (8, 6, 5, 4, 3),
while holding the other components at their baseline precision of 32 bits.
Figures~\ref{fig:group-quantization-accVSBitops} (a,b,c,g,h,i) report accuracy vs BitOps results for the six models described in Table~\ref{tab:models} when quantizing only one component of the KAN layer at a time, i.e., $W^{(l)}$, $A^{(l)}$, and $B^{(l)}$.
As shown in these figures, the weight matrix $W^{(l)}$ is the most sensitive to quantization, consistent with previous findings~\cite{QuantKAN}. 
Concerning the activation matrix $A^{(l)}$, we observe it is less sensitive overall than $W^{(l)}$ (with the exception of LeKAN), but contributes less to BitOps reduction, as anticipated in Section~\ref{sec:approach}.
Finally, we characterize, for the first time, the quantization sensitivity of $B^{(l)}$, the intermediate tensor of B-spline activations, which proves quite robust.
For instance, the CNN3 model, trained on CIFAR-10, has a sharp accuracy drop of $\approx 12\%$ for $W^{(l)}$ quantized to 5 bits, whereas it only drops by $2\%$ for $A^{(l)}$ at the same bit-width and practically shows no degradation for $B^{(l)}$ quantized to 5 and even 4 bits.
More generally, all models maintain nearly full accuracy even when $B^{(l)}$ is quantized to 3 or 4 bits, with CNN3 being the most sensitive ($\approx 3.5\%$ drop at 3 bits). Moreover, quantizing $B^{(l)}$ also greatly contributes to BitOps reduction.
%The magnitude of the BitOps reduction obtained when quantizing $B^{(l)}$ varies significantly across architectures. 
The CNN and ResKAN models exhibit the largest relative savings. For instance, quantizing $B^{(l)}$ to $3$-bits in the ResKAN18 model reduces BitOps from $\approx4 \times 10^9$ to $\approx6 \times 10^8$, i.e., almost one order of magnitude reduction. The CNN4 model shows a similarly strong drop in BitOps from $4.83 \times 10^9$ to $6.5 \times 10^8$.

Then, we quantize all the tensor components of the KAN layer, $W^{(l)}$, $A^{(l)}$, and $B^{(l)}$, at the same time, and explore all combinatorial bit-width reduction possibilities. In Figure~\ref{fig:group-quantization-accVSBitops} (d,e,f,j,k,l), we report the obtained Pareto-front in terms of Accuracy vs BitOps trade-off, i.e., the best quantization combinations of $W^{(l)}$, $A^{(l)}$, and $B^{(l)}$. For easier comparison, we align the x-axis values across graphs showing the same model (e.g., a and d).
As expected, across all tested models, the combined quantization yields a consistent reduction in BitOPs compared to quantizing the KAN components one at a time, without loss of accuracy. For example, for ResKAN18, while quantizing the single $B^{(l)}$ component allowed reducing the BitOPs by almost an order of magnitude (from $\approx4\times10^9$ to $\approx6\times10^8$) without accuracy loss (Figure~\ref{fig:group-quantization-accVSBitops}(i)), the combined quantization allowed further reduction of more than 60\% ($\approx2\times10^8$) without accuracy loss (Figure~\ref{fig:group-quantization-accVSBitops}(l)). 
In general, we observe that solutions providing both high accuracy and BitOps reductions typically use $W^{(l)}$ quantization with 5-8 bits (i.e., red, purple, and pink colors in the graphs), while $B^{(l)}$ quantization values are most of the time as low as 3 bits (i.e., pentagon shape in the graphs), and $A^{(l)}$ quantization values typically range from 5 to 8 bits with some models going as low as 4 bits (e.g., KANMLP1 KANMLP2 and LeKAN). $W^{(l)}$ quantization values lower than 5 (i.e., green and orange colors) mostly entail substantial accuracy drops, as also happens for $A^{(l)}$ quantization values lower than 4 or 5.

% Figure~\ref{fig:group-quantization-memVSBitops} shows the same combined quantization from a memory perspective, reporting the Pareto front of accuracy versus total memory (weights + activations) for each model.
% \MT{comment the figure~\ref{fig:group-quantization-memVSBitops}}
% \input{memory-figure.tex}

\subsection{B-Spline Tabulation Space Exploration}
\label{sec:results-B-spline}
In this subsection, we explore the space of possible B-Spline tabulation opportunities and show the effects in terms of accuracy vs memory tradeoff (Sec. \ref{sec:results-accuracyvsmemory}), accuracy vs computational complexity (Sec. \ref{sec:results_bspline_complexity}), inference latency on GPUs and a systolic-array-based hardware accelerator, i.e., KAN-SAs~\cite{KANSAS} (Sec. \ref{sec:results_bspline_latency}), and area and frequency scaling of the latter (Sec. \ref{sec:results_bspline_hw}).

\subsubsection{Accuracy vs Memory tradeoff}
\label{sec:results-accuracyvsmemory}

\begin{figure*}[htbp]
    \centering
    \begin{subfigure}[b]{0.32\textwidth}
        \includegraphics[width=\textwidth]{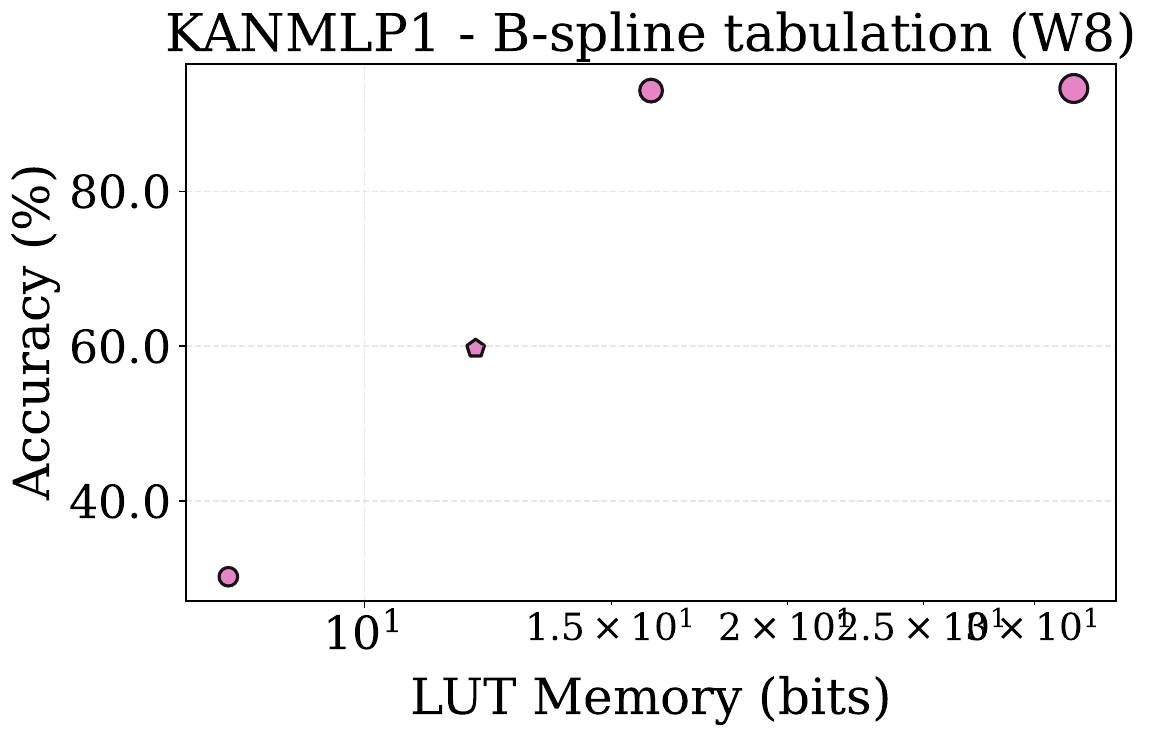}
        \vskip -7pt \caption{}
    \end{subfigure}
    \begin{subfigure}[b]{0.32\textwidth}
        \includegraphics[width=\textwidth]{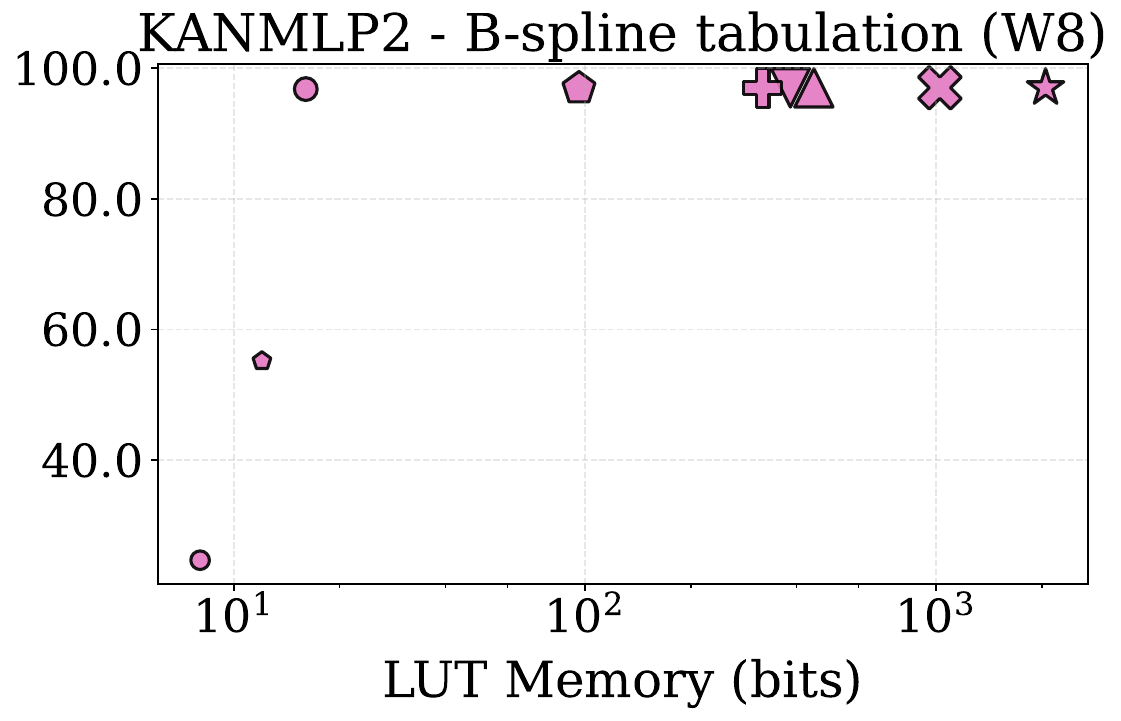}
        \vskip -7pt \caption{}
    \end{subfigure}
    \begin{subfigure}[b]{0.32\textwidth}
        \includegraphics[width=\textwidth]{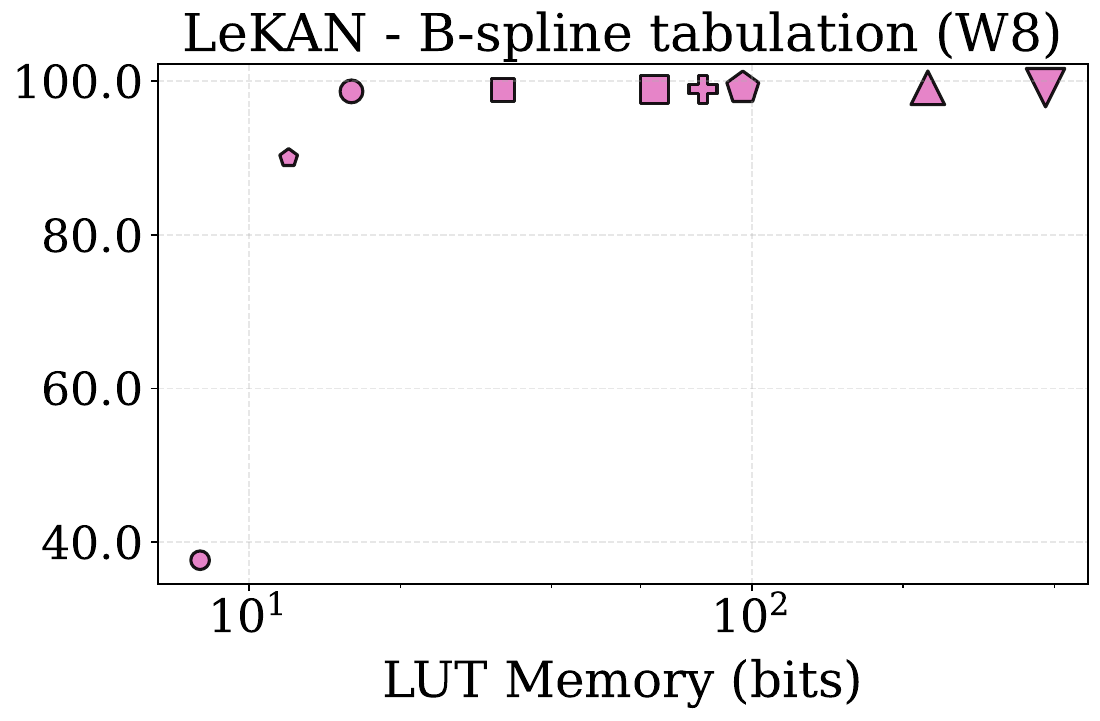}
        \vskip -7pt \caption{}
    \end{subfigure}
    \begin{subfigure}[b]{0.32\textwidth}
        \includegraphics[width=\textwidth]{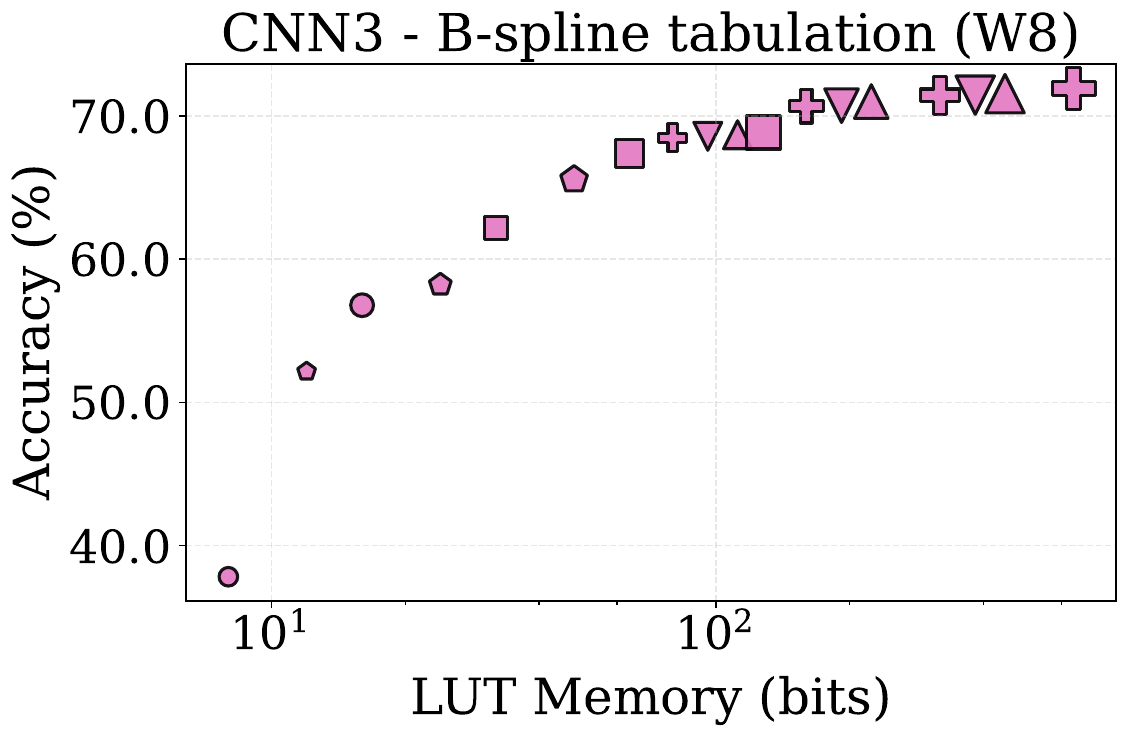}
        \vskip -7pt \caption{}
    \end{subfigure}
    \begin{subfigure}[b]{0.32\textwidth}
        \includegraphics[width=\textwidth]{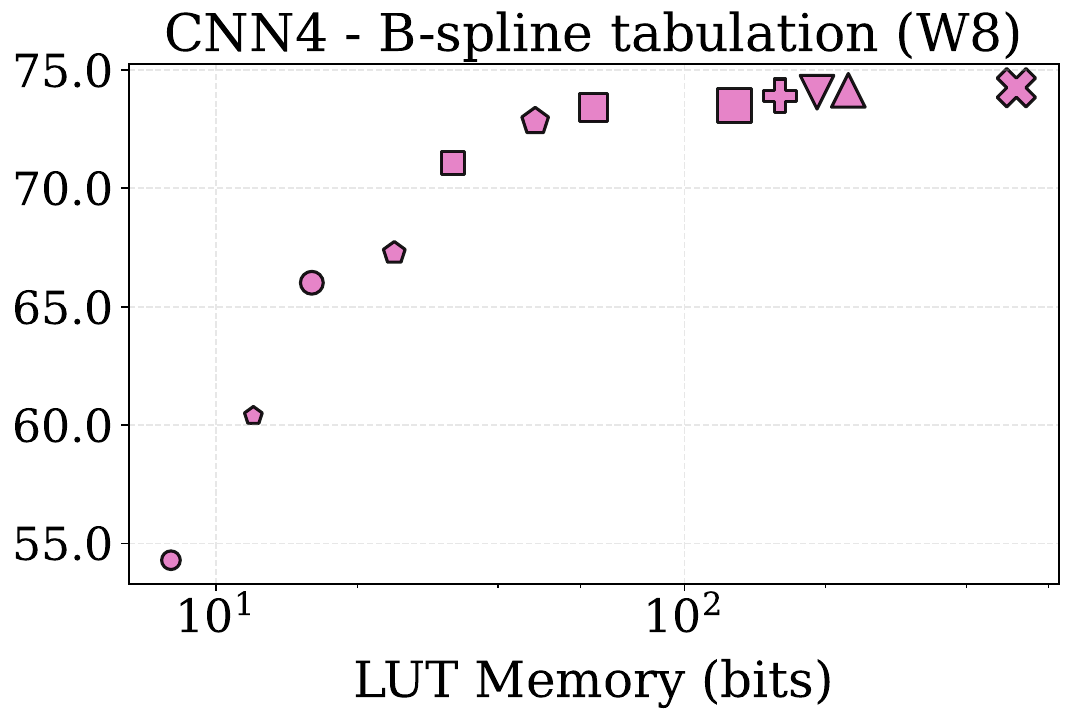}
        \vskip -7pt \caption{}
    \end{subfigure}
    \begin{subfigure}[b]{0.32\textwidth}
        \includegraphics[width=\textwidth]{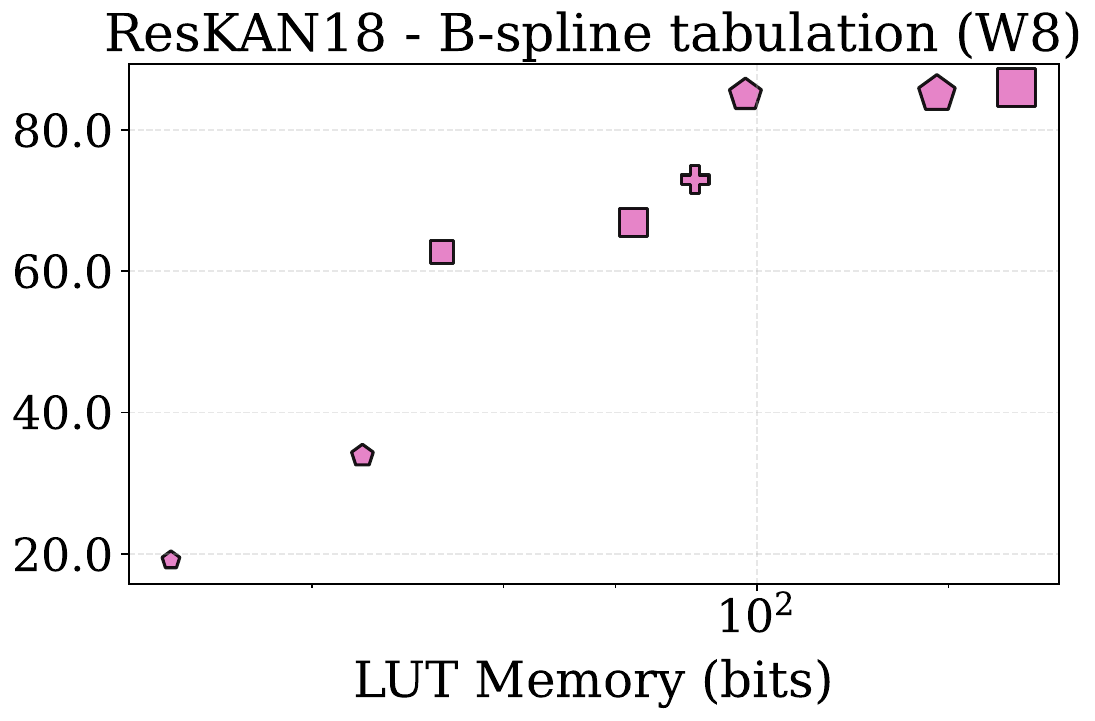}
        \vskip -7pt \caption{}
    \end{subfigure}
    \begin{subfigure}[c]{0.85\textwidth}
        \includegraphics[width=\textwidth]{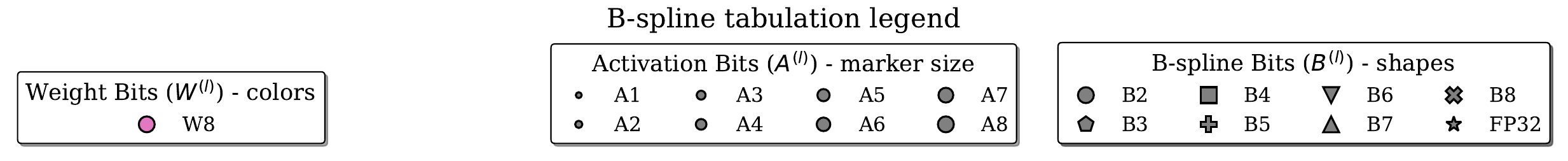}
    \end{subfigure}
    \caption{Accuracy vs LUT Memory trade-off for B-spline tabulation with 8-bit weights $W^{(l)}$.
    The graphs show the Pareto front of accuracy versus LUT memory cost for different bit-widths of $A^{(l)}$ (marker size) and stored values $B^{(l)}$ (marker shape).}
    \label{fig:lutmem-vs-acc}
    \vspace{-10pt}
\end{figure*}

In Figure~\ref{fig:lutmem-vs-acc}, we report the Pareto front of accuracy versus LUT memory for each model with fixed 8-bit weights $W^{(l)}$, varying the addressing bit-width $bw_{A^{(l)}}$ and stored value bit-width $bw_{B^{(l)}}$. Some plots do not report solutions with higher bit widths (e.g., KANMPL1) since they were Pareto-dominated by the reported low-bit-width ones.
For all models, $bw_{B^{(l)}}$ can be reduced to 3 or 4 bits with minimal accuracy loss, with associated LUT memory gains up to more than one order of magnitude when moving from 8 bits (X shape) to 3 (pentagon) or 2 (dot) bits, confirming the robustness of the B-spline activations already observed in Figure~\ref{fig:group-quantization-accVSBitops}. 
The addressing bit-width $bw_{A^{(l)}}$ has the largest impact on LUT memory due to the exponential scaling ($2^{(bw_{A^{(l)}})}$ entries), but also on accuracy: reducing $bw_{A^{(l)}}$ below 5 bits causes significant degradation for most models. The MNIST models %(KANMLP1, KANMLP2, LeKAN) 
are the most tolerant, maintaining near-baseline accuracy even at $bw_{A^{(l)}}\!=\!4$ with $bw_{B^{(l)}}\!=\!3$, while the convolutional models %(CNN3, CNN4, ResKAN18) 
degrade faster at low $bw_{A^{(l)}}$ values.

\subsubsection{Accuracy vs Computational Complexity}
\label{sec:results_bspline_complexity}

\begin{figure*}[htbp]
    \centering
    \begin{subfigure}[b]{0.32\textwidth}
        \includegraphics[width=\textwidth]{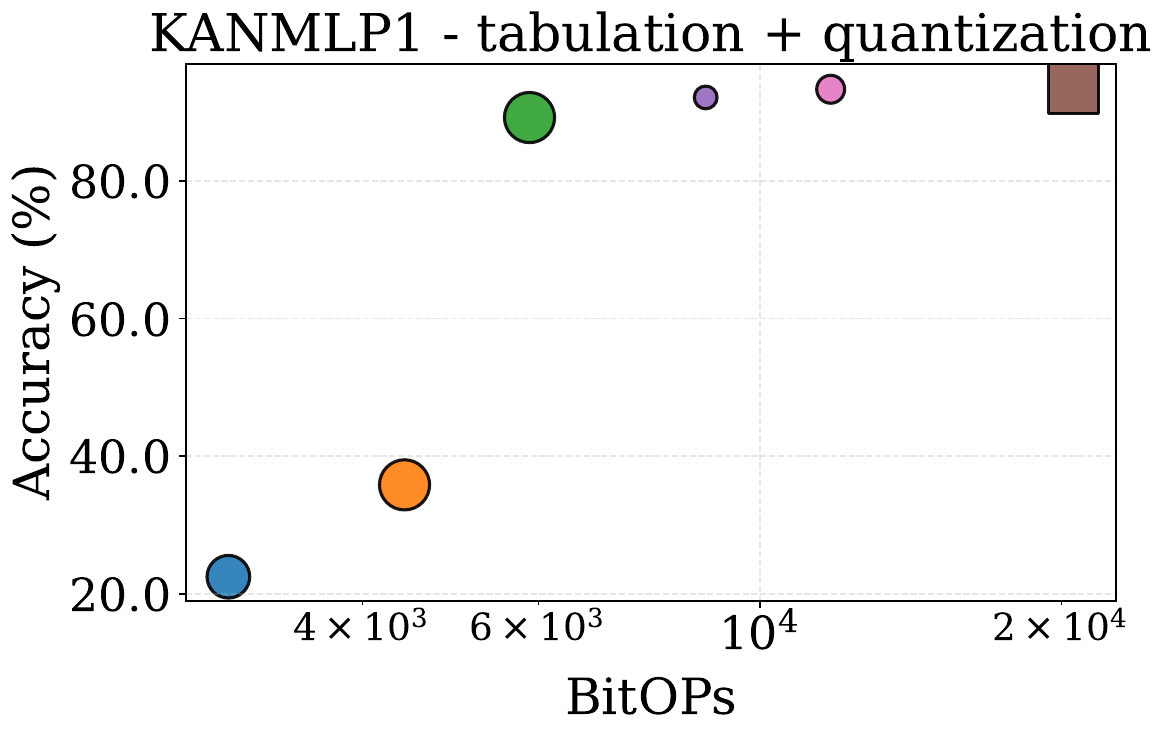}
        \vskip -7pt \caption{}
    \end{subfigure}
    \begin{subfigure}[b]{0.32\textwidth}
        \includegraphics[width=\textwidth]{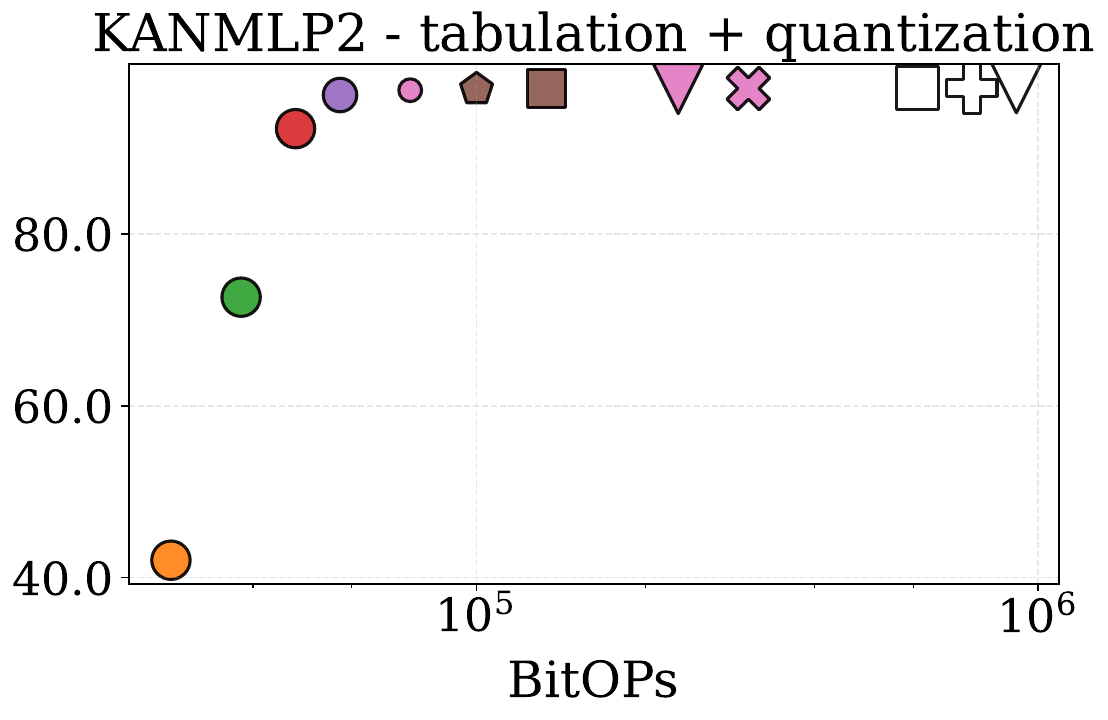}
        \vskip -7pt \caption{}
    \end{subfigure}
    \begin{subfigure}[b]{0.32\textwidth}
        \includegraphics[width=\textwidth]{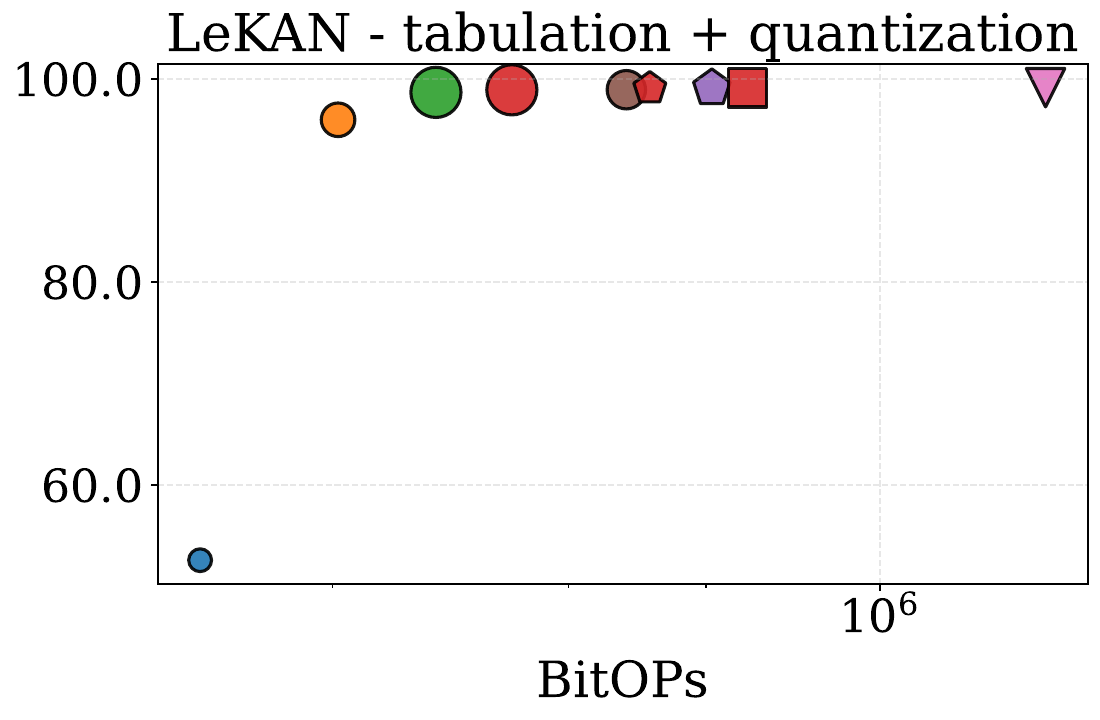}
        \vskip -7pt \caption{}
    \end{subfigure}
    \begin{subfigure}[b]{0.32\textwidth}
        \includegraphics[width=\textwidth]{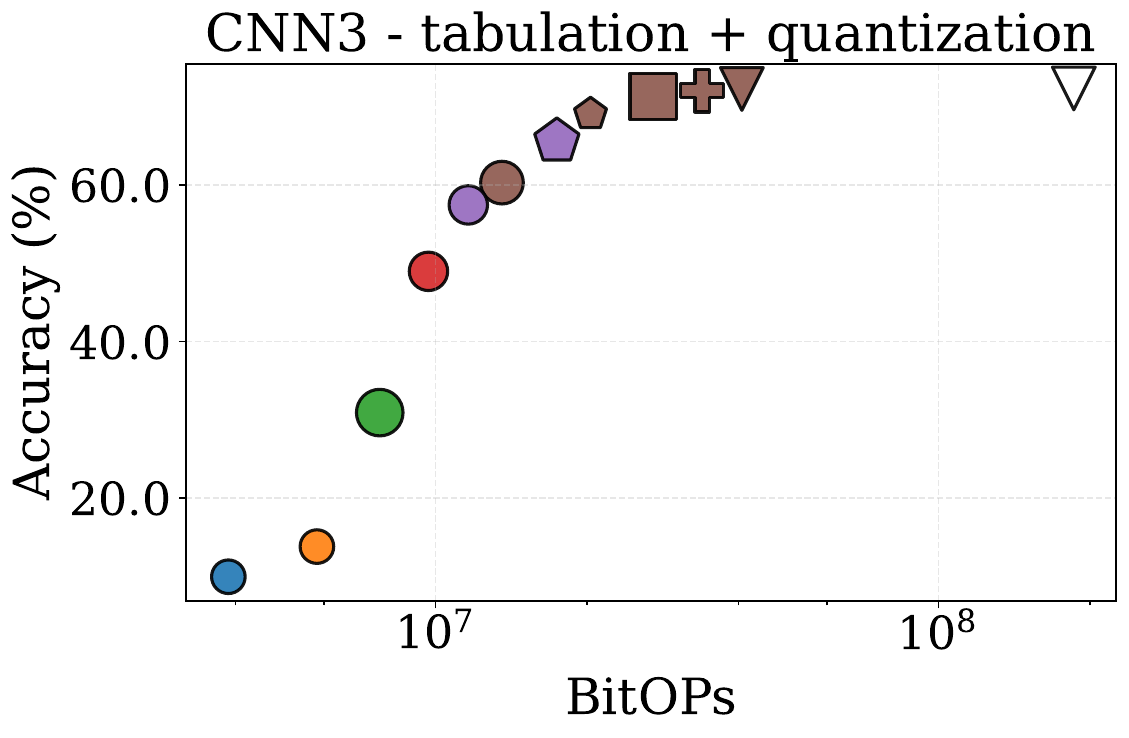}
        \vskip -7pt \caption{}
    \end{subfigure}
    \begin{subfigure}[b]{0.32\textwidth}
        \includegraphics[width=\textwidth]{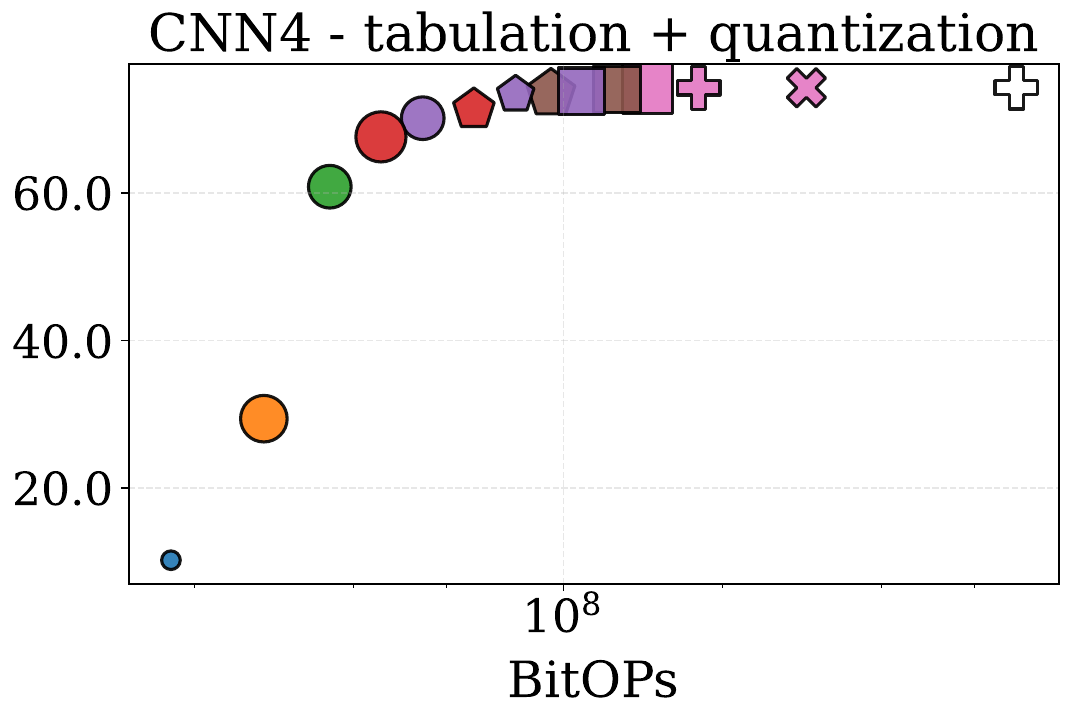}
        \vskip -7pt \caption{}
    \end{subfigure}
    \begin{subfigure}[b]{0.32\textwidth}
        \includegraphics[width=\textwidth]{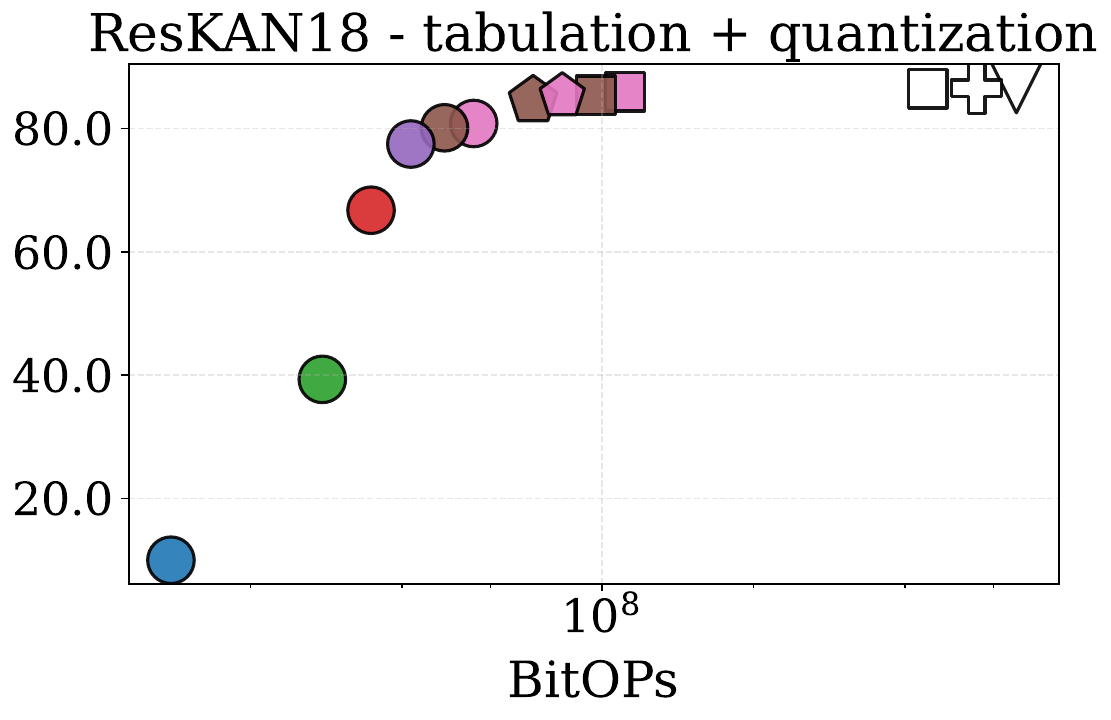}
        \vskip -7pt \caption{}
    \end{subfigure}
    \begin{subfigure}[c]{0.85\textwidth}
        \includegraphics[width=\textwidth]{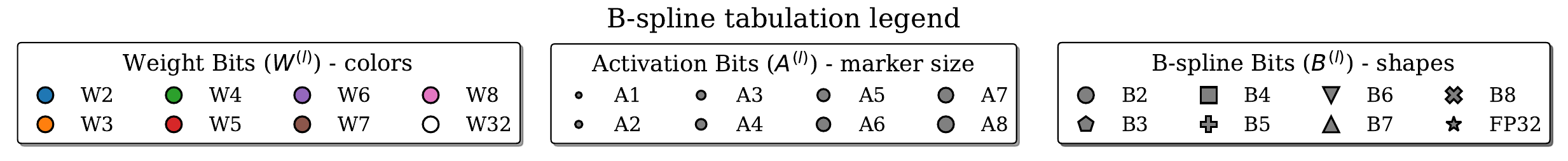}
    \end{subfigure}
    \caption{Accuracy vs BitOPs trade-off for B-spline tabulation combined with quantization.
    By replacing Cox-de Boor evaluation with a LUT, the B-spline computational cost is eliminated, leaving only the matrix multiplication.
    The bit-width of $W^{(l)}$ is encoded by marker color, that of $A^{(l)}$ (LUT addressing) by marker size, and that of $B^{(l)}$ (LUT stored values) by marker shape.}
    \label{fig:wlutq-bitops}
    \vspace{-10pt}
\end{figure*}

As mentioned, B-spline tabulation also further reduces computational complexity and runtime.
Concerning computational complexity, in Figure~\ref{fig:wlutq-bitops}, we report the Pareto front of accuracy versus BitOPs when combining B-spline tabulation with quantization. Compared to Figure~\ref{fig:group-quantization-accVSBitops}, the elimination of the Cox-de Boor evaluation shifts the Pareto front to significantly lower BitOPs, as the only remaining computational cost is the weight matrix multiplication. For instance, in Figure~\ref{fig:group-quantization-accVSBitops}(l) for ResKAN18, it was possible to reduce BitOPs by one order of magnitude (from $\approx7 \times 10^8$ to $\approx7 \times 10^7$) without accuracy loss.
Considering both quantization and tabulation, the total BitOPs reduction from the FP32 baseline is substantial. For ResKAN18, from $\approx4 \times 10^9$ in Figure~\ref{fig:group-quantization-accVSBitops}(i) to $\approx7 \times 10^7$ in Figure~\ref{fig:wlutq-bitops}(l), i.e., more than $50\times$ reduction without accuracy loss.

\subsubsection{Latency Improvement}
\label{sec:results_bspline_latency}

Finally, in Table~\ref{tab:forwardlatency}, we report the inference speedup on GPU obtained thanks to B-spline tabulation compared with the direct evaluation using the recursive Cox-de Boor formula (Eq.~\ref{eq:deboor2}).
As shown in the table, B-spline tabulation consistently accelerates inference across different grid size $G$ values and model architectures, from single- and two-layer KAN MLPs for MNIST classification, to ConvKAN models based on LeNet and deeper CNNs,
and even ResNet-based ConvKAN. Tabulation yields $2.11$--$2.87\times$ total speedup for $G\!=\!3$ and up to $3.36\times$ for $G\!=\!5$.
%
% Also, we observe that the B-spline evaluation time depends on the model's depth and layer widths, which affect B-spline evaluation and lookup operations in opposite ways.
% Indeed, the B-spline evaluation, based on the Cox-de Boor recursion, requires multiple sequential kernel executions per
% layer, one for each recursion step. Since GPUs can efficiently parallelize the element-wise operations across
% the width of each layer, increasing layer width has a limited impact on the total B-spline time.
% In contrast, increasing model depth multiplies the number of sequential kernel launches, making depth the primary cause of B-spline slowdown. This explains why ResKAN18 (20 layers) has a significantly higher B-spline time than CNN3 or CNN4 (3, 4 layers), despite the latter having wider layers.
% In any case, tabulating B-splines consistently improves inference time by more than $2\times$. \MT{moved this text here from section C, as it discusses more table III than VII. Actually I'm not sure I entirely get the text, let's discuss.}

However, GPUs only support specific bit-widths natively (e.g., 8-bit integer) and do not benefit from further reducing the B-spline table precision below 8 bits.
Custom hardware accelerators such as KAN-SAs~\cite{KANSAS} can leverage lower-bit B-spline tables. Reducing the precision of the stored values shrinks each processing element (PE), allowing a larger systolic array to fit within the same area budget.
\begin{table}[!t]
\caption{GPU time (ms) on RTX3090 for the full test set (10 000 samples) with 8-bit B-spline tabulation (256 entries), measured with \texttt{torch.profiler}. $P\!=\!3$ for all models.}
\centering
\small
\begin{tabular}{l c r r r r}
\hline
Model & $G$ & Baseline & BSP \% & BSP Tab. & Speedup \\
      &     & (ms)     &        & (ms)     &         \\
\hline
\multicolumn{6}{l}{\textit{MNIST}} \\
\multirow{2}{*}{KANMLP1}  & $3$ & $13.9$ & $96$ & $4.8$   & $2.87\times$ \\
                           & $5$ & $17.6$ & $97$ & $5.2$   & $3.36\times$ \\
\multirow{2}{*}{KANMLP2}  & $3$ & $15.2$ & $97$ & $5.4$   & $2.83\times$ \\
                           & $5$ & $19.3$ & $98$ & $5.9$   & $3.29\times$ \\
\multirow{2}{*}{LeKAN}    & $3$ & $763.3$ & $81$ & $359.8$ & $2.12\times$ \\
                           & $5$ & $923.9$ & $84$ & $373.3$ & $2.47\times$ \\
\hline
\multicolumn{6}{l}{\textit{CIFAR-10}} \\
\multirow{2}{*}{CNN3}     & $3$ & $769.8$ & $80$ & $355.8$ & $2.16\times$ \\
                           & $5$ & $932.3$ & $83$ & $373.0$ & $2.50\times$ \\
\multirow{2}{*}{CNN4}     & $3$ & $898.7$ & $78$ & $426.5$ & $2.11\times$ \\
                           & $5$ & $1086.2$ & $81$ & $446.5$ & $2.43\times$ \\
\multirow{2}{*}{ResKAN18} & $3$ & $6499.3$ & $81$ & $2977.4$ & $2.18\times$ \\
                           & $5$ & $7969.4$ & $84$ & $3184.2$ & $2.50\times$ \\
\hline
\end{tabular}
\label{tab:forwardlatency}
\end{table}
To quantify this effect on custom hardware, we synthesize a TPUv1-like~\cite{tpuv1} accelerator using a KAN-SAs-style weight-stationary systolic array~\cite{KANSAS} as its compute core on an SQRL Acorn CLE-215+ board featuring a Xilinx Artix-7 xc7a200t FPGA.
All configurations use 8-bit weights and activations, $G\!=\!5$, $P\!=\!3$, and 256-entry B-spline lookup tables; only the B-spline table value bit-width varies.
In the KAN-SAs architecture, each PE stores a local copy of the B-spline lookup table. Thus, reducing the bit-width of the stored values shrinks each PE, freeing FPGA slices for additional PEs.
\begin{table}[!b]
\caption{FPGA resource utilization and average measured inference time over 5 MNIST-shaped MLP workloads $[784, n, 10]$ with $n \in \{32, 64, 128, 192, 256\}$ (batch\,=\,10\,000) with varying B-spline table bit-width ($G\!=\!5$, $P\!=\!3$, 50~MHz).}
\centering
\setlength{\tabcolsep}{2.5pt}
\begin{tabular}{c c r@{/}l r@{/}l c c}
\hline
B & Array & \multicolumn{2}{c}{Slice LUTs} & \multicolumn{2}{c}{Slices} & CC & Time \\
(bits) & size & \multicolumn{2}{c}{(\%)} & \multicolumn{2}{c}{(\%)} & & (ms) \\
\hline
8 & $15 \times 15$ & 113.7k & 85\% & 32.1k & 96\% & 5 206k & 183.0 \\
7 & $16 \times 16$ & 124.9k & 93\% & 33.3k & 100\% & 4 219k & 156.4 \\
5 & $18 \times 18$ & 119.7k & 89\% & 32.8k & 98\% & 3 618k & 162.1 \\
4 & $19 \times 19$ & 114.4k & 86\% & 33.2k & 99\% & 3 286k & 157.9 \\
3 & $20 \times 20$ & 115.2k & 86\% & 32.6k & 98\% & 2 969k & 147.0 \\
\hline
\end{tabular}
\label{tab:fpga_bspwidth}
\end{table}
% FPGA measured accuracy (15x15 50MHz unless noted): bsp8=95.74%, bsp7_16x16=95.74%, bsp5_18x18=95.69%, bsp4_19x19=95.68%, bsp3_20x20=95.59%, bsp3_70MHz=95.59%
%
For each B-spline bit-width, the largest square systolic array that still fits within the FPGA's slice budget is selected (i.e., increasing the array dimensions by one beyond those reported causes the design to exceed the available slices and fail place-and-route).
Note that in this TPUv1-like configuration, weight preload and matrix computation execute sequentially: the systolic array must be fully loaded with the current weight tile before input data can be streamed through. For single-sample inference, this per-tile preload overhead grows with the array dimensions and can offset the benefit of having more PEs, particularly when the layer dimensions are small relative to the array.
However, for batched inferences, the preload cost is amortized across all samples in the batch, making larger arrays strictly beneficial.
For each bit-width (B), Table~\ref{tab:fpga_bspwidth} reports the largest array dimension achieved, the FPGA resource utilization (Slice LUTs and Slices), the compute cycles (CC), and average measured inference time over 5 MNIST-shaped MLP workloads $[784, n, 10]$ with $n \in \{32, 64, 128, 192, 256\}$ on the full test set (10 000 samples).
As shown, reducing the B-spline precision from 8 to 7 bits allows scaling the systolic array from $15 \times 15$ to $16 \times 16$, yielding a measured $1.17\times$ average speedup.
Reducing the precision further to 3 bits enables a $20 \times 20$ array, achieving a measured $1.24\times$ average speedup over the 8-bit baseline.
The gap between estimated and measured speedup for larger arrays is due to DMA transfer overhead, which grows with the array dimensions due to memory bus alignment constraints.

\begin{table}[!b]
\caption{Maximum clock frequency for a $16 \times 16$ KAN-SAs systolic array with varying B-spline table bit-width on FPGA ($W\!=\!8$, $G\!=\!5$, $P\!=\!3$).}
\centering
\setlength{\tabcolsep}{2.5pt}
\begin{tabular}{c r@{\,/\,}l c}
\hline
B & \multicolumn{2}{c}{Slice LUTs} & Max Freq \\
(bits) & \multicolumn{2}{c}{(\%)} & (MHz) \\
\hline
8$^\dagger$ & \multicolumn{2}{c}{---} & --- \\
7 & 124.9k & 93\% & 50 \\
6 & 118.1k & 88\% & 60 \\
5 & 97.9k & 73\% & 70 \\
4 & 85.2k & 64\% & 75 \\
3 & 79.1k & 59\% & 75 \\
\hline
\multicolumn{3}{l}{{\scriptsize $^\dagger$exceeds available slices}}
\end{tabular}
\label{tab:fpga_fmax}
% \vspace{-13pt}
\end{table}

Beyond throughput, the area savings from B-spline quantization also enable higher clock frequencies.
To isolate this effect, Table~\ref{tab:fpga_fmax} reports the maximum clock frequency achievable for a fixed $16 \times 16$ array with each B-spline bit-width.
The 8-bit configuration cannot complete place-and-route for a $16 \times 16$ array.
Reducing the bit-width to 7 bits allows the design to fit, but routing congestion at 93\% LUT utilization limits the clock to 50~MHz.
Further quantization progressively reduces utilization and relieves routing pressure, enabling higher clock frequencies: 60~MHz at 6 bits, 70~MHz at 5 bits, and 75~MHz at 3 and 4 bits, a 50\% improvement over the 7-bit baseline.
Alternatively, the freed resources could be used for architectural enhancements such as double-buffered weight preload to overlap tiling overhead with computation, or deeper on-chip buffers to hide memory latency.
\begin{figure*}[b]
  %\vspace{-5pt}
    \centering
    \begin{subfigure}[b]{0.32\textwidth}
        \includegraphics[width=\textwidth]{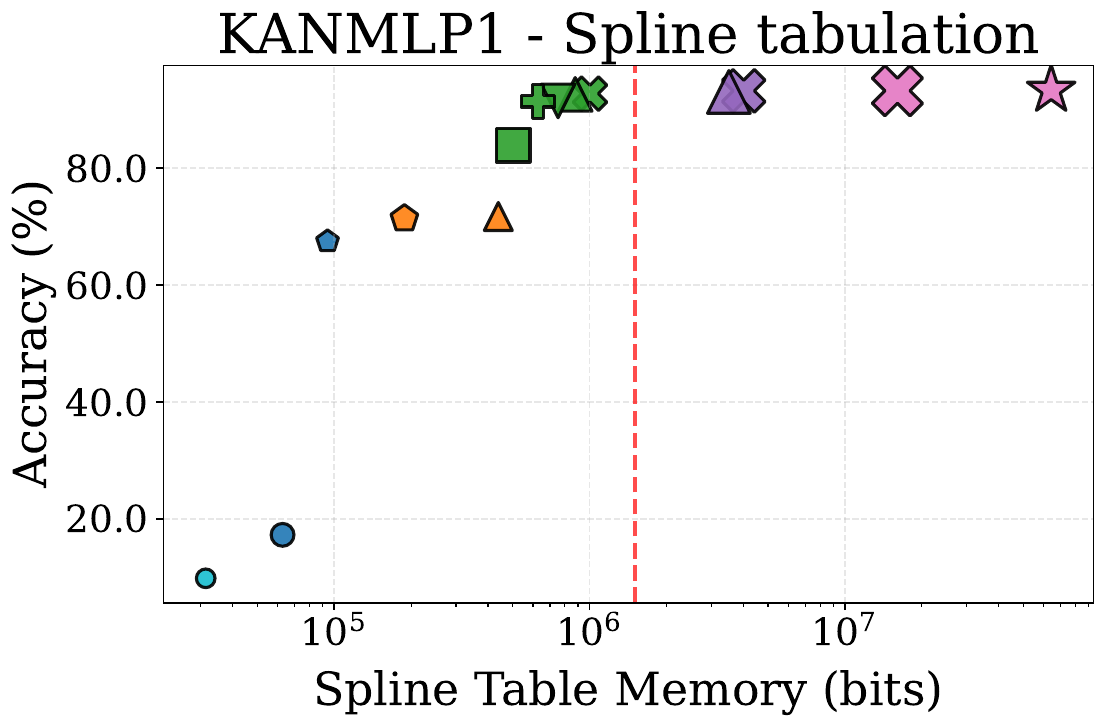}
        \vskip -7pt \caption{}
    \end{subfigure}
    \begin{subfigure}[b]{0.32\textwidth}
        \includegraphics[width=\textwidth]{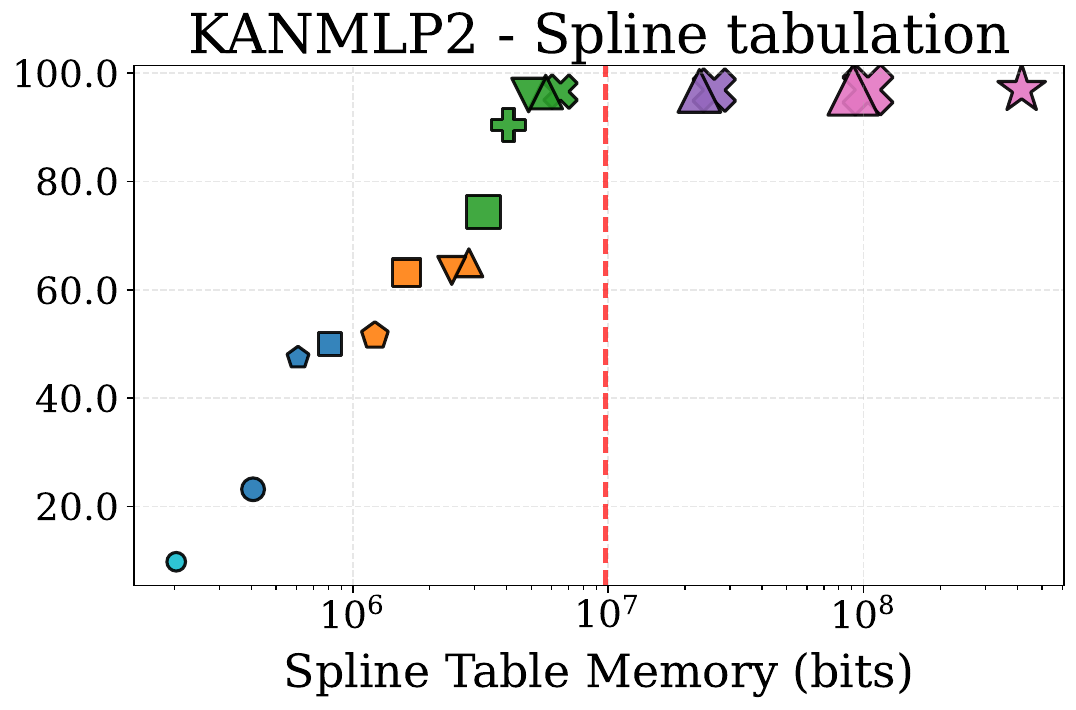}
        \vskip -7pt \caption{}
    \end{subfigure}
    \begin{subfigure}[b]{0.32\textwidth}
        \includegraphics[width=\textwidth]{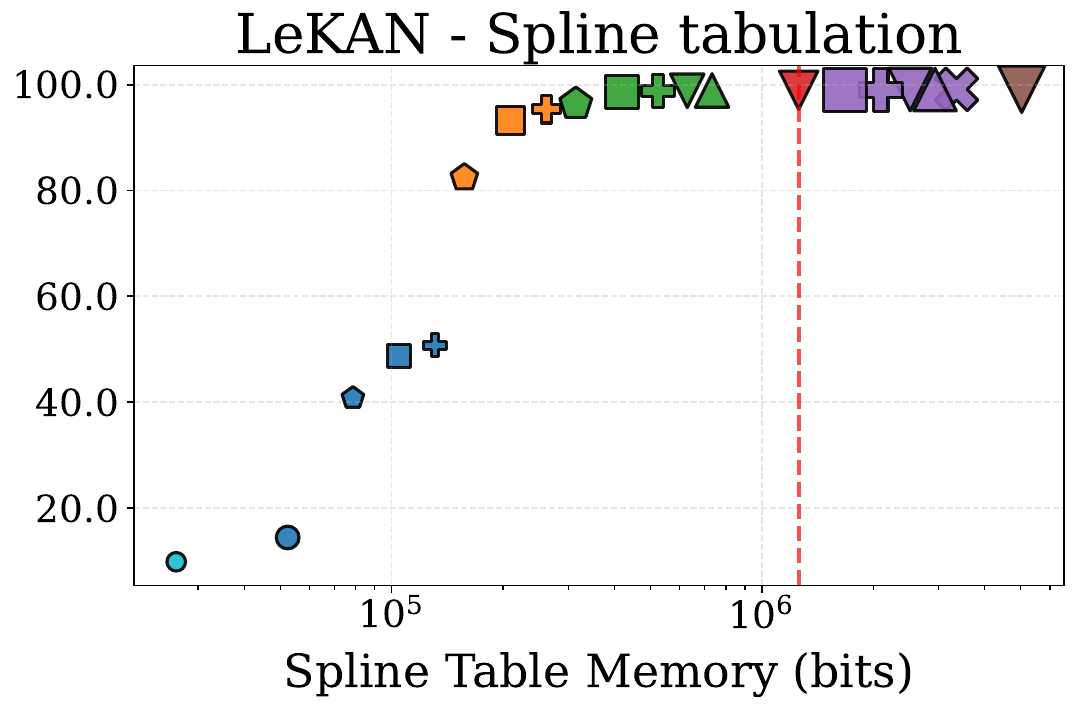}
        \vskip -7pt \caption{}
    \end{subfigure}
    \begin{subfigure}[b]{0.32\textwidth}
        \includegraphics[width=\textwidth]{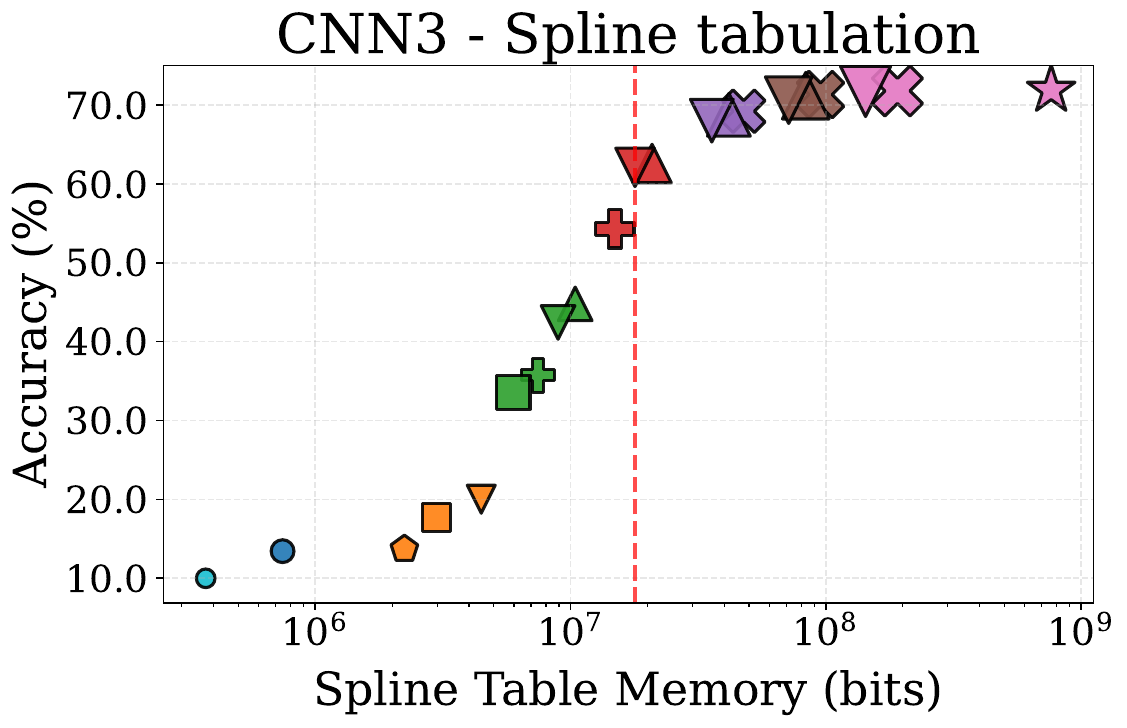}
        \vskip -7pt \caption{}
    \end{subfigure}
    \begin{subfigure}[b]{0.32\textwidth}
        \includegraphics[width=\textwidth]{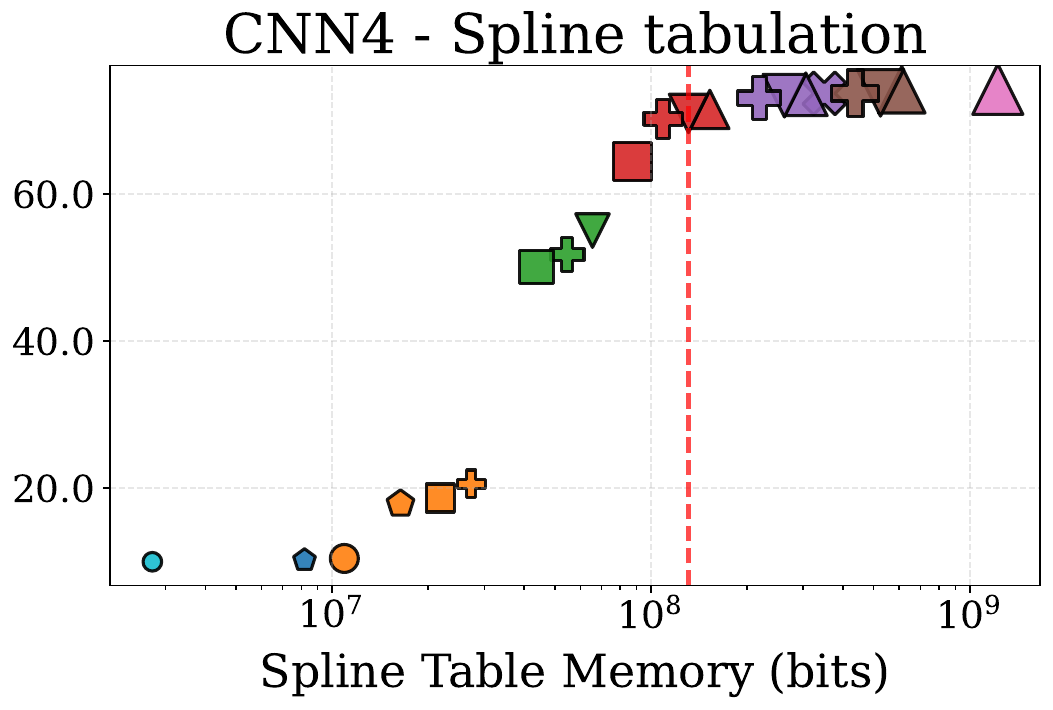}
        \vskip -7pt \caption{}
    \end{subfigure}
    \begin{subfigure}[b]{0.32\textwidth}
        \includegraphics[width=\textwidth]{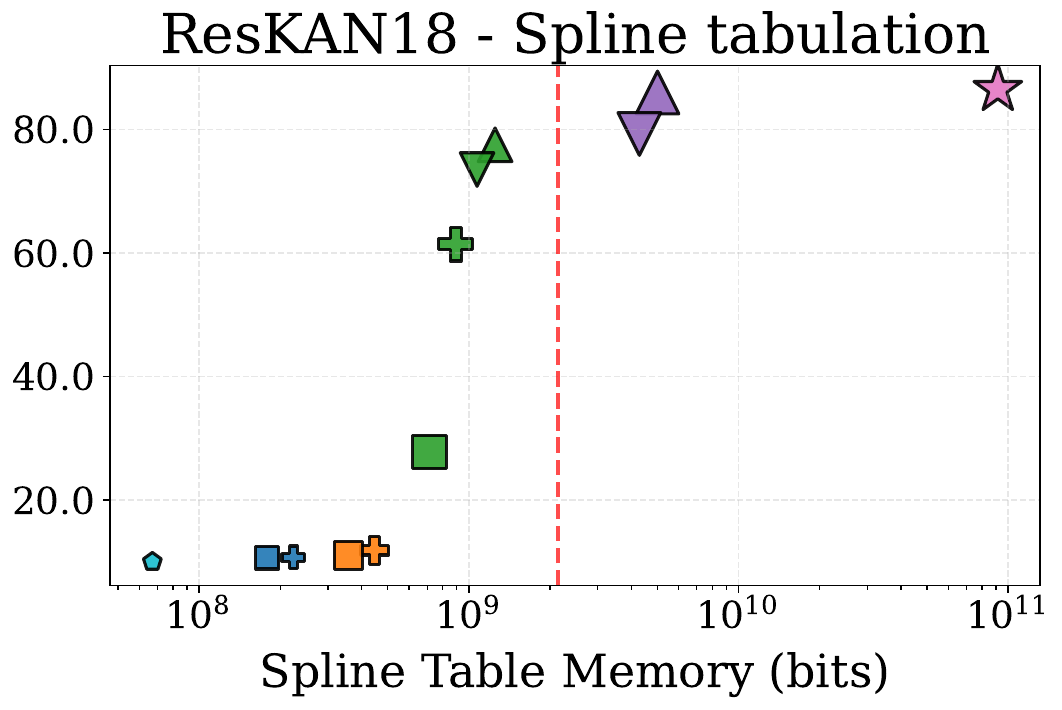}
        \vskip -7pt \caption{}
    \end{subfigure}
    \begin{subfigure}[c]{0.85\textwidth}
        \includegraphics[width=\textwidth]{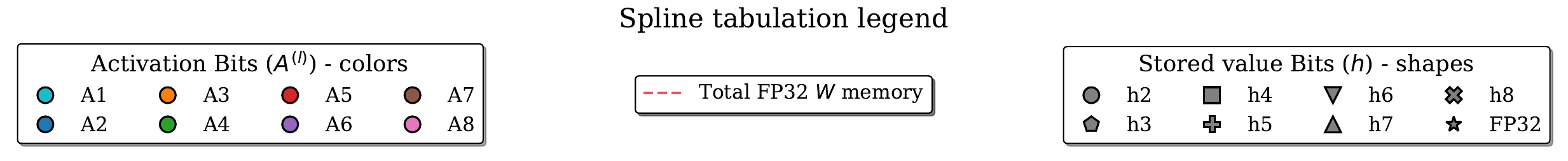}
    \end{subfigure}
    \caption{Accuracy vs Spline Table Memory trade-off for spline tabulation.
    The graphs show the Pareto front of accuracy versus total spline table memory cost for different bit-widths of $A^{(l)}$ (marker size) and stored value precision $h$ (marker shape).}
    \label{fig:splmem-vs-acc}
\end{figure*}

To validate this, we run the 3-bit $16 \times 16$ array at 75~MHz and measure the inference time for the KAN MLP $[784, 64, 10]$ on the full MNIST test set.
Compared to the 7-bit $16 \times 16$ baseline at 50~MHz, the measured time drops from $185$~ms to $129.55$~ms, achieving a $1.43\times$ speedup.
% The improvement is consistent across 10 random two-layer MLP workloads with varying layer dimensions, where the average inference time decreases from $32.67$~ms to $22.82$~ms ($\mathbf{1.43\times}$).
Note that the achievable frequency remains limited by the high fanout of the control 
signals across the systolic array, as the KAN-SAs design targets ASIC and is not optimized for FPGA.

\subsubsection{KAN-SAs Area and Frequency Scaling}
\label{sec:results_bspline_hw}

To evaluate the impact of B-spline quantization on hardware independently of system-level effects, we synthesize the KAN-SAs $16 \times 16$ systolic array in isolation using Synopsys Design Compiler targeting ST 28nm FD-SOI technology.
Table~\ref{tab:asic_bspwidth} reports the cell area and maximum achievable clock frequency for each configuration.
Halving both the B-spline and weight precision from 16 to 8 bits reduces the cell area by 60\% (from 1\,426k to 574k~$\mu m^2$ at 1~GHz) and raises the maximum frequency from 833 to 1\,111~MHz.
Further reducing only the B-spline precision from 8 to 3 bits yields an additional 30\% area reduction (to 403k~$\mu m^2$) and increases the maximum frequency to 1\,250~MHz, a 50\% advantage over the 16-bit baseline.
Overall, the combined reduction from $B\!=\!16$, $W\!=\!16$ to $B\!=\!3$, $W\!=\!8$ achieves a 72\% area saving.

\begin{table}[t]
\caption{ASIC synthesis results for a $16 \times 16$ KAN-SAs systolic array with varying B-spline and weight precision (ST 28nm FD-SOI, $G\!=\!5$, $P\!=\!3$).}
\centering
\setlength{\tabcolsep}{4pt}
\begin{tabular}{c c c c c}
\hline
B & W & Area @ 1~GHz & \multicolumn{2}{c}{Max frequency} \\
(bits) & (bits) & ($\mu m^2$) & (MHz) & Area ($\mu m^2$) \\
\hline
16 & 16 & 1 425 590$^\dagger$ & 833 & 1 479 594 \\
\hline
8 & 8 & 574 102 & 1 111 & 596 850 \\
7 & 8 & 541 978 & 1 176 & 581 353 \\
6 & 8 & 515 865 & 1 176 & 532 536 \\
5 & 8 & 470 712 & 1 250 & 489 225 \\
4 & 8 & 433 345 & 1 333 & 469 248 \\
3 & 8 & 403 158 & 1 250 & 434 962 \\
\hline
\multicolumn{5}{l}{\footnotesize $^\dagger$Timing violated at 1~GHz; area shown for comparison.}
\end{tabular}
\label{tab:asic_bspwidth}
\vspace{-10pt}
\end{table}

\subsection{Spline Tabulation space exploration} \label{sec:spline_tabul_results}

In this subsection, we explore the space of possible quantized \textit{Spline} tabulation opportunities and show the effects in terms of accuracy vs. memory trade-off (Sec. \ref{sec:results-spline-accuracyvsmemory}) and inference latency (Sec. \ref{sec:results_spline_latency}).

\subsubsection{Accuracy vs Memory tradeoff of Spline Tabulation}
\label{sec:results-spline-accuracyvsmemory}

Figure~\ref{fig:splmem-vs-acc} shows the Pareto front of accuracy versus total spline table memory for all models. The vertical dashed line indicates the total FP32 coefficient memory $\sum_l N_{in}^{(l)} \cdot N_{out}^{(l)} \cdot (G+P) \cdot 32$ bits.
%
%\SE{@MT, I changed this paragraph a bit, and added colors to plot for clarity:}%MT: TOP!!!
%
Tabulating a whole KAN leads to $\sum_l N_{in}^{(l)} \cdot N_{out}^{(l)} \cdot 2^{(bw_{A^{(l)}})} \cdot h$  stored bits when using $h$-bit quantized spline values and $2^{(bw_{A^{(l)}})}$ table entries, as already mentioned in Section~\ref{sec:spllut} for one layer.
For the MNIST models (KANMLP1, KANMLP2, LeKAN), several Pareto-optimal configurations lie to the left of this line with less than 1\% accuracy drop, meaning the spline tables require less storage than the coefficients they replace. Specifically, $bw_{A^{(l)}}$ can be reduced to as few as 4 bits and $h$ to 6 bits with negligible accuracy loss, consistent with prior studies~\cite{kanele}.
For CIFAR-10, the sensitivity is model-dependent. CNN3, the smallest CIFAR-10 model, requires $bw_{A^{(l)}} \geq 7$ bits to stay within ${\sim}1\%$ of full precision, while the deeper ResKAN18 is more sensitive to $h$, needing $h \geq 7$ bits. Overall, under per-tensor post-training quantization alone, the table memory required to maintain accuracy exceeds the original coefficient storage by a wide margin, making spline tabulation impractical for larger models without additional compression.

% \subsubsection{Accuracy vs Computational Complexity }
% \label{sec:results-spline-accuracyvscomplexity}

\subsubsection{Latency Improvement}
\label{sec:results_spline_latency}

\begin{table}[b]
\caption{GPU time (ms) on RTX3090 for spline tabulation with 256 entries and 32-bit values, measured with \texttt{torch.profiler}. $P\!=\!3$ for all models.}
\centering
\small
\begin{tabular}{l c r r r}
\hline
Model & $G$ & Baseline & Spline Tab. & Speedup \\
      &     & (ms)     & (ms)        &         \\
\hline
\multicolumn{5}{l}{\textit{MNIST}} \\
\multirow{2}{*}{KANMLP1}  & $3$ & $13.9$   & $1.9$   & $7.3\times$ \\
                           & $5$ & $17.6$   & $1.9$   & $9.1\times$ \\
\multirow{2}{*}{KANMLP2}  & $3$ & $15.2$   & $6.4$   & $2.4\times$ \\
                           & $5$ & $19.3$   & $6.4$   & $3.0\times$ \\
\multirow{2}{*}{LeKAN}    & $3$ & $763.3$  & $201.0$ & $3.8\times$ \\
                           & $5$ & $923.9$  & $200.8$ & $4.6\times$ \\
\hline
\multicolumn{5}{l}{\textit{CIFAR-10}} \\
\multirow{2}{*}{CNN3}     & $3$ & $769.8$  & $1436.5$ & $0.5\times$ \\
                           & $5$ & $932.3$  & $1448.0$ & $0.6\times$ \\
\multirow{2}{*}{CNN4}     & $3$ & $898.7$  & $6293.9$ & $0.1\times$ \\
                           & $5$ & $1086.2$ & $6517.4$ & $0.2\times$ \\
\multirow{2}{*}{ResKAN18} & $3$ & $6499.3$ & $5672.5$ & $1.1\times$ \\
                           & $5$ & $7969.4$ & $5692.3$ & $1.4\times$ \\
\hline
\end{tabular}
\label{tab:spltab_latency}
\end{table}

\begin{figure}[t]
\centering
\includegraphics[width=.9\columnwidth, trim={10 3 5 5}, clip]{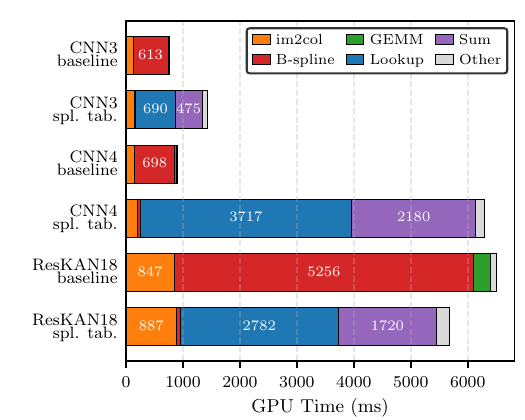}
\caption{CUDA kernel time breakdown for baseline (Cox-de Boor) vs spline tabulation on CNN3, CNN4, and ResKAN18 
($G\!=\!3$, $P\!=\!3$, RTX3090).}
\label{fig:kernel_gantt}
\end{figure}

Table~\ref{tab:spltab_latency} reports the inference latency of spline tabulation compared to the baseline recursive B-spline-based execution.
As discussed in Section~\ref{sec:spllut}, spline tabulation replaces the entire KAN layer with $N_{in} \cdot N_{out}$ table lookups from the precomputed spline tables, followed by a summation along $N_{in}$ to produce each output, eliminating both the Cox-de Boor recursion and the coefficient matrix multiply.
Since each lookup contributes a single addition, the operation is inherently memory-bound, unlike the matrix multiply, which benefits from highly optimized GEMM implementations.
Thus, spline tabulation yields a net speedup only if the time saved by avoiding B-spline evaluation exceeds the cost of using memory-bound lookups instead of efficient matrix multiplication.
For small models (KANMLP1, KANMLP2, LeKAN), \textit{spline} tabulation yields speedups of $2.4$--$9.1\times$ on GPU, as shown in Table~\ref{tab:spltab_latency}. Indeed, the \textit{spline} tables remain reasonably small (see Figure~\ref{fig:splmem-vs-acc}), and, in the baseline version, the B-spline dominates the whole computation (see Table~\ref{tab:forwardlatency}, BSP\% column).
However, for larger models (CNN3, CNN4), the speedup drops significantly (down to $0.5\times$ for CNN3 and $0.1\times$ for CNN4), while ResKAN18 achieves a $1.4\times$ speedup.
By analyzing this phenomenon, we observe that speedups depend on model depth and width: the baseline's B-spline recursive evaluation is sequential and scales poorly with depth (requiring multiple kernel runs per layer). Conversely, increasing the layer width has little impact on its computation time, since GPUs can efficiently parallelize the element-wise operations across the width of each layer.
Concerning the ConvKAN layer's spline tables, their indexing performs $K^2\cdot C_\text{in} \cdot C_\text{out} \cdot H_\text{out} \cdot W_\text{out}$ scattered memory reads. A deep model distributes reads across many independent kernel runs while maintaining good memory locality, whereas increasing layer width degrades memory throughput by concentrating lookups into a single kernel.
Therefore, ResKAN18 requires more B-spline time due to its deeper architecture ($20$ layers), despite narrower layers, as confirmed in Figure~\ref{fig:kernel_gantt}. CNN4's final layer (resolution $8\times8$) requires $64$ accesses per table per sample, whereas ResKAN18's smaller layers ($2\times2$) require only $4$ accesses.
In summary, \textit{spline tabulation} is slower than the baseline recursive B-spline for CNN3 and CNN4 because the cumulative lookup cost exceeds the savings from reduced recursion.

\subsubsection{Scalability of Spline Tabulation on FPGA}
\label{sec:results_spline_scalability}

\begin{figure}[b]
  \centering
  \includegraphics[width=.95\columnwidth]{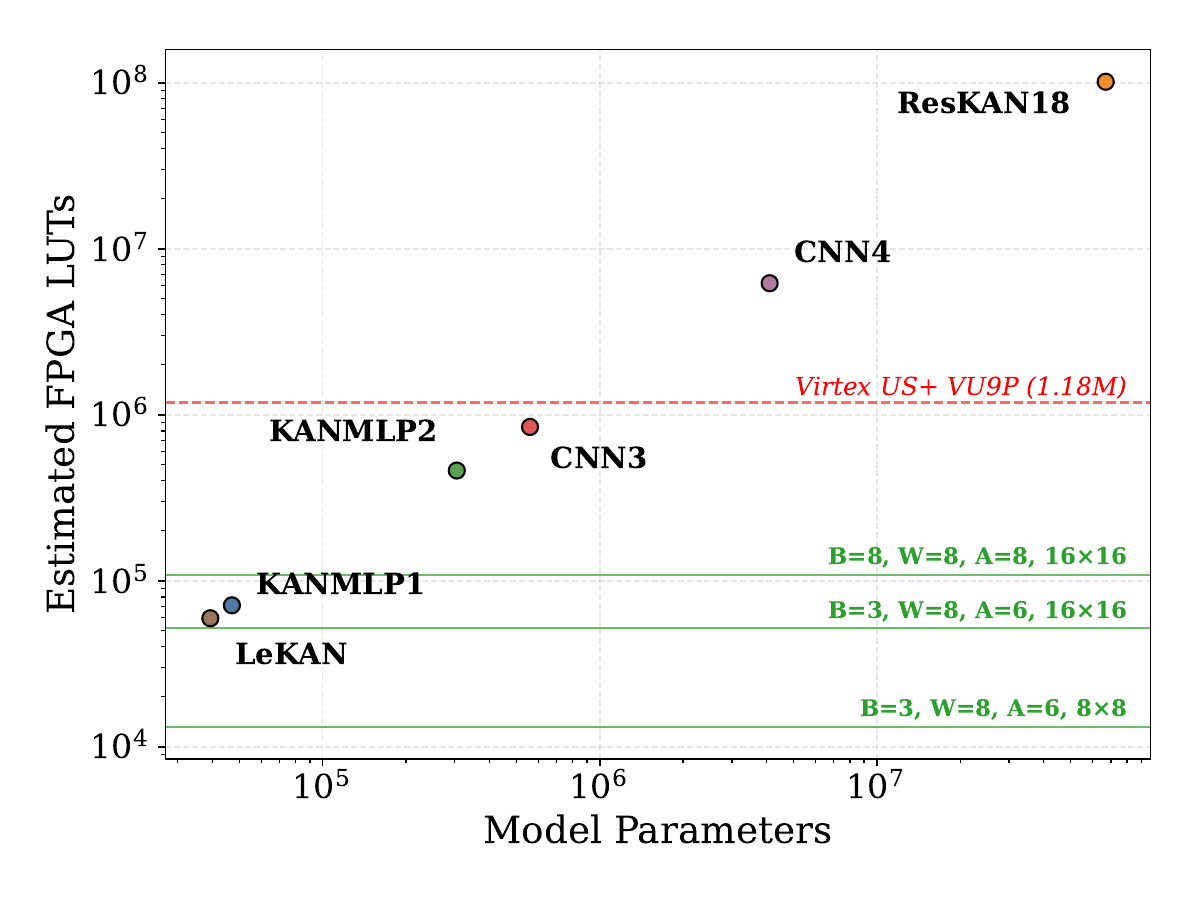}
  \caption{Estimated FPGA LUT cost of spline tabulation~\cite{kanele} for the models studied in this paper. The dashed red line indicates the capacity of a Virtex UltraScale+ FPGA. Horizontal green lines report KAN-SAs systolic array~\cite{KANSAS} configurations, labeled by B-spline coefficient precision~$B$, weight precision~$W$, activation precision~$A$, and array dimensions.}
  \label{fig:kanele_scaling}
\end{figure}

The $N_{in} \cdot N_{out}$ scaling of spline tabulation, which limits GPU speedups for larger models (as shown in Table~\ref{tab:spltab_latency}), also poses a challenge for FPGA implementations. In an approach such as~\cite{kanele}, each spline connection requires independent FPGA LUTs, leading to a total resource cost proportional to $\sum_l N_{in}^{(l)} \cdot N_{out}^{(l)}$.
The work~\cite{kanele} explored spline tabulation implementations on various FPGAs, including mid-range and high-end devices such as Zynq UltraScale+ and Virtex UltraScale+.
Figure~\ref{fig:kanele_scaling} shows the estimated FPGA LUT cost for the models studied in this paper, using an empirical cost of approximately $9$ FPGA LUTs per connection derived from the synthesis results reported in the ablation study in~\cite{kanele} for 6-bit addressed, 8-bit valued spline tables on a Xilinx Virtex UltraScale+, a high-end FPGA family.
The graph shows that the spline tabulation approach remains viable for models of limited dimensions, such as KANMLP1, KANMLP2, LeKAN, and CNN3. However, it does not scale well for larger models, such as CNN4 and ResKAN18, exceeding the available resources by orders of magnitude on a high-end FPGA.
Using aggressive pruning (${\sim}98\%$) combined with quantization-aware training via Brevitas~\cite{brevitas}, the work~\cite{kanele} manages to reduce the cost of a KANMLP2 variant to $3809$ LUTs on a Zynq UltraScale+, albeit with 1-bit input addressing and only after pruning and re-training.
Whether such extreme pruning is feasible on more complex tasks and datasets remains to be determined, thus leaving open questions about the scalability of the spline tabulation approach.
In contrast, the KAN-SAs-like systolic array~\cite{KANSAS} accelerators have a constant area footprint regardless of model size, as shown by the horizontal lines in Figure~\ref{fig:kanele_scaling}.

\section{Conclusion}

Kolmogorov-Arnold Networks (KANs) have attracted much attention for their promise of better parameter efficiency and interpretability than Multi-Layer Perceptrons (MLPs). Nevertheless, their reliance on evaluating spline functions, often expressed as linear combinations of weighted basis splines (B-splines), poses a significant computational challenge. 
In this paper, we investigated the impact of low-bit quantization on KANs' accuracy as well as its benefits in reducing computational complexity and improving hardware efficiency. Our experiments revealed that B-splines are highly robust to quantization, i.e., down to 2-3 bits without loss of accuracy, and yield substantial reductions in computational complexity, whereas the learnable weights are more sensitive to quantization.
Moreover, the computational overhead introduced by B-splines recursive evaluation, can be reduced through B-spline tabulation. Hence, we evaluate the impact of quantization on tabulation and their joint benefits for computational complexity and hardware efficiency.
Our experiments showed that a BitOps reduction of more than $50\times$ is possible, e.g., for ResKAN18, without accuracy loss thanks to quantized B-spline tabulation. Moreover, we obtained up to $2.9\times$ inference speedup on GPUs thanks to B-spline tabulation, and up to 36\% resource reduction, 50\% higher clock frequency and $1.24\times$ inference speedup on FPGA thanks to quantized B-spline tabulation. On 28nm FD-SOI ASIC, a 72\% area reduction and 50\% higher maximum frequency was obtained. Finally, we explored the opportunities and limitations of a similar state-of-the-art approach tabulating the entire \textit{splines} instead of the B-splines, highlighting its convenience compared to the B-spline tabulation for models of limited dimension, whereas its scalability becomes challenging for larger models.

\section*{Acknowledgments}
This work was supported by the French National Research Agency (ANR) through the RADYAL project ANR-23-IAS3-0002.

\bibliography{refs.bib}
\bibliographystyle{IEEEtran}

\end{document}